\documentclass[article,floats,floatfix,amssymb,prd,onecolumn,superscriptaddress,nofootinbib]{revtex4-1}

\usepackage[english]{babel}
\usepackage[utf8]{inputenc}
\usepackage{amsmath}
\usepackage{mathbbol}
\usepackage{amssymb}
\usepackage{bbold}
\usepackage{graphicx,amsfonts}
\usepackage{float}
\usepackage{epsfig}
\usepackage[colorlinks=true,
linkcolor=blue,
urlcolor=blue,
citecolor=blue]{hyperref}
\usepackage{bm}
\usepackage{mathrsfs}
\usepackage{mathtools}
\usepackage{enumerate}
\usepackage{amsthm}
\usepackage{bbm}
\usepackage{comment}
\usepackage{physics}
\usepackage{url}
\usepackage{upgreek}
\usepackage{enumitem}
\usepackage[svgnames]{xcolor}
\usepackage[title]{appendix}
\newcommand{\sapienza}{Dipartimento di Fisica, Sapienza Università 
	di Roma, Piazzale Aldo Moro 5, 00185, Roma, Italy}
\newcommand{\infn}{INFN, Sezione di Roma, Piazzale Aldo Moro 2, 00185, Roma, Italy}
\newcommand{\PI}{Perimeter Institute for Theoretical Physics, 31 Caroline St N, Waterloo, ON N2L 2Y5, Canada}
\begin{document}

\author{Romeo Felice Rosato}
\email{romeofelice.rosato@uniroma1.it}
\affiliation{\sapienza}
\affiliation{\infn}

\author{Marina De Amicis}
\email{mdeamicis@perimeterinstitute.ca}
\affiliation{\PI}

\author{Paolo Pani}
\email{paolo.pani@uniroma1.it}
\affiliation{\sapienza}
\affiliation{\infn}

\date{\today}

\title{Singular structures and causality of the Schwarzschild Green's function\\ in the frequency domain}

\begin{abstract}
We study two singular spectral components of the Green's function of a Schwarzschild black hole and their interpretation in the frequency domain: 
(i) the low-frequency branch cut, which yields corrections to Price's law tails in the form of inverse power laws weighted by logarithmic terms; and 
(ii) the quasinormal-mode spectrum, which generates a redshifted response for sources extended toward the horizon. 
We show that the frequency-domain Green's function can be naturally interpreted in terms of greybody factors, providing the first analytical justification for recent phenomenological ringdown models based on these quantities.
For sources localized outside the peak of the potential barrier, we identify two tail contributions activated with a time delay, arising from backscattering of the prompt response and of the ringdown signal. 
We show that corrections to Price's law can be relevant at intermediate times, when the ringdown still dominates the waveform. 
For sources localized inside the potential barrier peak, the tail is suppressed and the signal is instead dominated by quasinormal frequencies. 
In this regime, these spectral components produce both the ordinary quasinormal-mode ringdown and an infinite tower of exponentially decaying terms governed by the horizon surface gravity, the so-called redshift terms. 
We demonstrate that this component is not screened by geometric features of the background spacetime and persists up to late times, as supported by numerical investigations of perturbative waveforms.
Our results provide a mathematical foundation for phenomenological modeling of the branch-cut contribution at intermediate times,
which is relevant for prospective observations of tails, and strong evidence for the presence of redshifted components from intermediate to late times.
\end{abstract}


\maketitle 

\twocolumngrid
\tableofcontents
\onecolumngrid


\section{Introduction}
The LIGO--Virgo--KAGRA network is building an expanding catalog of gravitational-wave detections from compact-object mergers~\cite{LIGOScientific:2016aoc,LIGOScientific:2021sio,LIGOScientific:2025hdt}, a trend that will continue with future detector upgrades and planned observatories~\cite{Saleem:2021iwi,ET:2025xjr,LISA:2024hlh}. These systems provide access to the strong-field regime of gravity, enabling precision tests of general relativity~\cite{Berti:2015itd,Berti:2018vdi,Cardoso:2019rvt,Berti:2025hly}.

The evolution of such systems consists of three stages: (i) the inspiral, during which the two objects orbit each other while gradually losing energy; (ii) the dynamical merger, when a single, highly distorted black hole is formed; and (iii) the ringdown, the final phase in which the remnant relaxes to a stable, stationary configuration, emitting a superposition of exponentially damped vibrations eventually leaving behind a non-oscillatory slow decaying signal, denoted as tail (see~\cite{Berti:2025hly} for a recent review).

The post-merger stage can be well described within black hole perturbation theory, providing a framework to test general relativity and its theoretical predictions.
In this context, the Green’s function of the perturbative problem is a powerful modeling tool.
The first systematic study of the Schwarzschild Green's function was carried out by Leaver~\cite{Leaver:1986gd}, who identified three distinct contributions: the prompt response, corresponding to the initial signal propagating along the light cone (see also~\cite{Andersson:1996cm}); the quasinormal mode (QNM) ringing, consisting of exponentially damped sinusoids; and a late-time power-law tail, first identified in~\cite{Price:1971fb}.

Despite substantial progress, a closed-form expression for the Green's function remains elusive, motivating continued efforts to deepen our understanding~\cite{DeAmicis:2025xuh,DeAmicis:2026tus,Su:2026fvj}.

When analytically continued to the complex frequency plane, the Schwarzschild Green's function exhibits three fundamental singular structures. These structures determine the time-domain response of the black hole and are therefore essential for its theoretical understanding. They consist of
\begin{itemize}
    \item a pole at zero frequency, $\omega = 0$, superposed with a branch point. This structure has recently been shown to account for the prompt response of the signal, both analytically~\cite{DeAmicis:2026tus} and numerically~\cite{Su:2026fvj};
    \item an infinite tower of complex, simple poles corresponding to the QNMs~\cite{Leaver:1985ax}. These modes dominate the Green's function at intermediate times, giving rise to exponentially damped oscillations in the signal;
    \item a branch cut along the negative imaginary axis, responsible for the late-time tail.
\end{itemize}

In this work, we analyze the last two of these structures, and provide a causal interpretation of the Green's function in the frequency domain.
We first study the branch-cut contribution for sources located outside the light ring. The branch cut is known to generate the late-time tail. Following its original prediction~\cite{Price:1971fb,Price:1972pw}, considerable effort has been devoted to understanding its late-time behavior~\cite{Cunningham:1978zfa,Cunningham:1979px,Leaver:1986vnb,Gomez:1992,Gundlach:1993tp,Gundlach:1993tn,Ching:1994bd,Burko:1997tb,Barack:1998bw,Bernuzzi:2008rq,Hod:2009my,Poisson:2002jz}. 
While the leading-order late-time tail is known to be universal for a broad class of systems~\cite{Rosato:2025rtr}, here we compute subleading contributions for an observer at future null infinity\footnote{Subleading corrections evaluated at future time-like infinity ($i^+$) were derived in Ref.~\cite{Casals:2015nja}. The analysis at $\mathscr{I}^+$ presented here corresponds to a physically distinct asymptotic regime.}, $\mathscr{I}^+$. 
We show that these contributions are not negligible in the reconstruction of the Green's function and play a significant role during the ringdown phase. While the fundamental mode and the lowest overtones remain dominant, the corrected tail attains amplitudes comparable to, and sometimes larger than, those of higher overtones.
This is relevant for potential observations of tails: the first measurable signal may arise at intermediate times after the merger rather than at late times, when the signal-to-noise ratio is significantly reduced.
We further show that beyond leading order the tail is no longer a simple power law, but instead involves a mixture of inverse powers of time multiplied by logarithmic corrections.  A public implementation of the computation is available at 
\href{https://github.com/romeofelicerosato-prog/tails_higherorders}{\texttt{tails\_higherorders}}.
We also analyze the causal structure of the late-time Green's function, demonstrating that two distinct tails are present for sources outside the light ring. One is generated by the back-scattering of the signal traveling directly from the source to the observer (also called prompt response), with the background curvature. 
The second comes from the backscattering of the signal that interacts with the potential barrier peak, the ringdown.
This identification extends to late-time tails the causal structure previously identified for QNMs in Ref.~\cite{DeAmicis:2025xuh} and for the prompt response in Ref.~\cite{DeAmicis:2026tus}.

Our causal decomposition of the Green's function in the frequency domain also offers a clear interpretation
of the signal from
sources outside the light ring 
in terms of greybody factors. Greybody factors have recently been proposed as a robust frequency-domain observable in ringdown, first for plunging particle scenarios~\cite{Oshita:2023cjz,Rosato:2024arw,Oshita:2024fzf,Rosato:2024arw} and subsequently for full binary black hole coalescences~\cite{Rosato:2025ulx,Okabayashi:2024qbz}. In particular, Ref.~\cite{Rosato:2025ulx} showed that the high-frequency portion of numerical-relativity simulations of coalescing binary black holes can be modeled with very small mismatches ($\mathcal{O}(10^{-5})$ for the $\ell=m=2$ mode) using the black hole reflection amplitude multiplied by a simple power-law function of the frequency. Here we provide the first analytical explanation for why the reflection amplitude is imprinted in the ringdown, thereby offering a theoretical justification for the phenomenological models introduced in those works. 

In the second part of this work we consider sources inside the light ring. We show that the late-time tail contribution from such sources is strongly suppressed due to tunneling through the gravitational effective potential barrier. We then focus on the QNM pole structure. By implementing the causality condition derived in Ref.~\cite{DeAmicis:2025xuh} in the frequency domain, we show that the modes are redshifted to new frequencies corresponding to so-called \emph{redshift terms}.
This contribution
is different from the \emph{horizon modes}, which have been the subject of an active debate~\cite{Dafermos:2005eh,Mino:2008at,Zimmerman:2011dx,Laeuger:2025zgb,DeAmicis:2025xuh,Oshita:2025qmn,Kankani:2026byb}. Some works~\cite{Mino:2008at,Zimmerman:2011dx} have suggested that these modes may appear as additional poles of the convolution integral between Green's function and test-particle source in the frequency domain, whereas Ref.~\cite{Oshita:2025qmn} argued that they are not actual poles, but are instead screened. Properly accounting for causality, here we show that, 
when considering a test-particle source inside the light ring, no new poles arise. However, redshifted contributions associated with the ordinary QNM poles appear, incidentally with the same frequency as the horizon modes. This confirms the results of Ref.~\cite{DeAmicis:2025xuh} through an independent frequency-domain analysis and corroborates that redshift terms persist  up until late times.  We also present a numerical indication for the presence of redshift terms in the case of a particle plunging inside the light ring. 

The paper is organized as follows. Sec.~\ref{sec:framework} introduces the theoretical framework. Sec.~\ref{sec:sourceoutside} discusses sources outside the light ring, with emphasis on late-time tails and greybody factors. Sec.~\ref{sec:sourceinside} focuses on sources inside the light ring, analyzing the redshift contributions to the Green's function and their connection to the greybody factor picture. 


\section{Framework}\label{sec:framework}
We study linear perturbations on a fixed Schwarzschild spacetime, whose line
element reads
\begin{equation}
ds^{2}=-\left(1-\frac{2M}{r}\right)dt^{2}
+\left(1-\frac{2M}{r}\right)^{-1}dr^{2}
+r^{2}\bigl(d\theta^{2}+\sin^{2}\!\theta\, d\phi^{2}\bigr)\, .
\end{equation}
Exploiting the spherical symmetry of the background, the metric perturbations can
be decomposed into spherical-harmonic modes, so that each $(\ell,m)$ component is
encoded in a single time--radial function $h_{\ell m}(t,r)$.

It is convenient to separate the perturbations into two independent sectors,
according to their transformation properties under parity.
Modes with parity $(-1)^{\ell}$ define the even (polar) sector, whereas those with
parity $(-1)^{\ell+1}$ belong to the odd (axial) one.
For each sector, a gauge-invariant master variable can be constructed: the
Regge--Wheeler function $\Psi^{o}_{\ell m}$ for odd modes and the Zerilli function
$\Psi^{e}_{\ell m}$ for even modes.
Both satisfy wave-like equations of the form
\begin{equation}
\left[\,\partial_{t}^{2}-\partial_{r_*}^{2}+V^{e/o}_{\ell}(r)\right]
\Psi^{e/o}_{\ell m}(t,r_*)=0\,, \label{Schroedinger}
\end{equation}
where $r_*$ is the tortoise coordinate, defined through
\begin{equation}
\frac{dr_*}{dr}=\left(1-\frac{2M}{r}\right)^{-1},
\end{equation}
and the effective potential $V^{e/o}_{\ell m}(r)$ depends on the parity sector.

At large radii, the master functions reproduce the gravitational-wave
polarizations according to
\begin{equation}
h_{\ell m}(t,r)=
\frac{1}{r}\sqrt{\frac{(\ell+2)!}{(\ell-2)!}}\,
\bigl(\Psi^{e}_{\ell m}+i\,\Psi^{o}_{\ell m}\bigr)
+{\cal O}(r^{-2})\,,
\end{equation}
up to the usual selection rules that eliminate the odd (resp., even)  contribution
when $\ell+m$ is even (resp., odd).

In the presence of a source, the perturbation equation takes the form
\begin{equation}
\left[\,\partial_{t}^{2}-\partial_{r_*}^{2}+V^{e/o}_{\ell}(r)\right]
\Psi^{e/o}_{\ell m}(t,r_*)
=\mathcal{S}^{e/o}_{\ell m}(t,r_*)\,,
\label{eq:RWZ_equation}
\end{equation}
where the source term may encode the initial data through 
\begin{equation}
\mathcal{S}^{\rm ID}_{\ell m}(t,r_*)=
\Psi_{\ell m}(t_{0},r_*)\,\partial_{t}\delta(t-t_{0})
+\partial_{t}\Psi_{\ell m}(t_{0},r_*)\,\delta(t-t_{0})\,,
\end{equation}
or arise from a non-vanishing energy--momentum tensor, as in the plunging particle
scenario~\cite{Zerilli:1970wzz}.
In this case, the source can be written as~\cite{Martel:2003jj,Sasaki:2003xr,Martel:2005ir,Nagar:2006xv}
\begin{equation}
S_{\ell m}(t,r_*)=f_{\ell m}(t,r_*)\delta(r_*-r_*(t))+g_{\ell m}(t,r_*)\partial_{r_*}\delta(r_*-r_*(t)) \, ,
\label{eq:testparticle_source}
\end{equation}
where $r_*(t)$ is the trajectory of the test-particle.
Explicit expressions for the functions $f_{\ell m}(t,r_*),\,g_{\ell m}(t,r_*)$ are given in Appendix~\ref{app:source}.

In the frequency domain, assuming a time dependence $\sim e^{-i\omega t}$, the response of a perturbed black hole obeys the
one-dimensional wave equation
\begin{equation}\label{eq:waveeq}
\left[\frac{d^2}{dr_*^2} + \omega^2 - V^{e/o}_{\ell}(r) \right]\tilde{\Psi}^{e/o}_{\ell m\omega}(r)
= \tilde{S}^{e/o}_{\ell m}(\omega, r)\,,
\end{equation}
where $\omega$ is the frequency. The source term $\tilde{S}_{\ell m}(\omega,r)$ encodes the perturbation mechanism. The homogeneous version of Eq.~\eqref{eq:waveeq} admits asymptotically, both at the horizon and at infinity, plane-wave solutions. 
We single out particular solutions according to their behaviour at the boundaries.
In particular, we define 
\begin{equation}\label{eq:hom_sol}
u^{\rm in}_{\ell m\omega} \to
\begin{cases}
A^{\rm in}_{\ell m\omega}\, e^{-i\omega r_*}
+ A^{\rm out}_{\ell m\omega}\, e^{i\omega r_*},
& r_* \to +\infty, \\[4pt]
e^{-i\omega r_*}, & r_* \to -\infty,
\end{cases}
\end{equation}
where $A^{\rm in/out}_{\ell m\omega}$ are the ingoing and outgoing wave amplitudes
at spatial infinity. These coefficients define the reflection and transmission
coefficients (also known as greybody factors)
\begin{equation}\label{eq:greybody}
\mathcal{R}_{\ell m\omega}
= \left|\frac{A^{\rm out}_{\ell m\omega}}{A^{\rm in}_{\ell m\omega}}\right|^2,
\qquad
\mathcal{T}_{\ell m\omega}
=1-\mathcal{R}_{\ell m\omega}= \left|\frac{1}{A^{\rm in}_{\ell m\omega}}\right|^2\, 
\end{equation}

A linearly independent solution 
is
\begin{equation}\label{eq:hom_sol2}
u^{\rm down}_{\ell m\omega}\to
\begin{cases}
e^{-i\omega r_*}, & r_* \to +\infty, \\[4pt]
B^{\rm in}_{\ell m\omega}\, e^{-i\omega r_*}
+ B^{\rm out}_{\ell m\omega}\, e^{i\omega r_*}, & r_* \to -\infty\,,
\end{cases}
\end{equation}
and the solution
$u^{\rm up}_{\ell m\omega}=u^{\rm down}_{\ell m,-\omega}$
provides a purely outgoing mode at infinity.

We define the Green's function as solution of Eq.~\eqref{eq:RWZ_equation}, with an impulsive source in $(t,r_*)$ coordinates, i.e.
\begin{equation}
\left[\,\partial_{t}^{2}-\partial_{r_*}^{2}+V^{e/o}_{\ell}(r)\right]
G_{\ell m}(t-t',r_*,r_*')
=\delta(t-t')\delta(r_*-r_*')\,,
\label{eq:GF_equation}
\end{equation}
The Green's function propagates information from the source to the observer, and must satisfy boundary conditions such that no information can exit the horizon nor enter from $\mathcal{I}^+$. This requirement implies that the frequency-domain Green's function can be written as 
\begin{equation}\label{eq:greenfunct}
\tilde{G}_{\ell m}(r_*,r'_*,\omega)
= \frac{i}{2 \omega A^{\rm in}_{\ell m\omega}}
\,\theta(r_*-r'_*)\,u^{\rm up}_{\ell m\omega}(r_*)\,
u^{\rm in}_{\ell m\omega}(r'_*)+\frac{i}{2 \omega A^{\rm in}_{\ell m\omega}}
\,\theta(r'_*-r_*)\,u^{\rm up}_{\ell m\omega}(r'_*)\,
u^{\rm in}_{\ell m\omega}(r_*)\,,
\end{equation}
This work will be restricted to compact sources localized between the black hole and the observer, so the second terms on the right-hand side will be discarded.
The time-domain retarded Green's function is then obtained anti-transforming Eq.~\eqref{eq:greenfunct}
\begin{equation}\label{eq:timegreenfunct}
G_{\ell m}(t-t',r_*,r_*')
= \frac{1}{2\pi}\int_{-\infty}^{+\infty} d\omega\,
\frac{i}{2 \omega A^{\rm in}_{\ell m\omega}}
u^{\rm up}_{\ell m\omega}(r_*)\,
u^{\rm in}_{\ell m\omega}(r'_*)\, e^{-i\omega(t-t')}\,.
\end{equation}

\begin{figure}[t]
    \centering
    \includegraphics[width=0.6\linewidth]{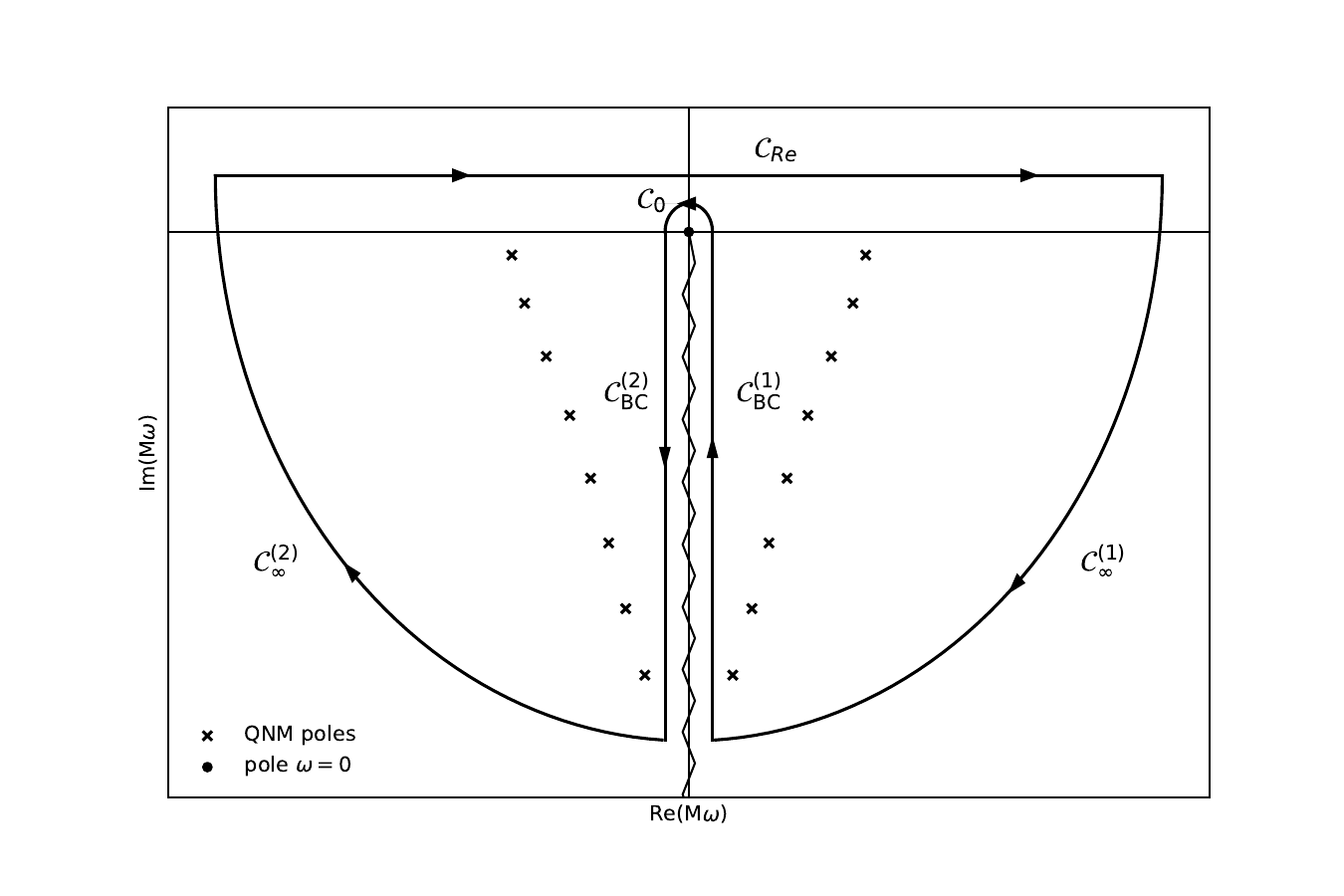}
    \caption{Schematic representation of the singular structure of the retarded Green's function in the complex $\omega$ plane. The branch cut along the negative imaginary axis, the zero-frequency singularity, and the QNM poles are indicated.}
    \label{fig:GF_singular_structure}
\end{figure}

As schematically illustrated in Fig.~\ref{fig:GF_singular_structure}, the full retarded Green's function in the frequency domain exhibits several singular structures in the complex $\omega$ plane~\cite{Leaver:1986gd}:
\begin{itemize}
    \item a branch cut extending along the negative imaginary axis;
    \item a singular structure at $\omega = 0$, where a pole merges with a branch point;
    \item a discrete set of poles corresponding to the QNMs.
\end{itemize}

Given the analytic continuation of the Green's function to the complex frequency plane, its time-domain representation can be obtained by deforming the integration contour as shown in Fig.~\ref{fig:GF_singular_structure}. By defining
\begin{equation}
    G_{\ell m}(t-t',r_*,r_*')={1 \over 2\pi}\int_{\mathcal{C}_{Re}} d\omega \,\tilde{G}_{\ell m}(\omega,r_*,r'_*)\, e^{-i \omega (t-t')}\,,
\end{equation}
and applying the residue theorem, together with the fact that the integral over the closed contour vanishes, and that the contribution from the arcs at infinity is zero by virtue of Jordan's lemma (see Ref.~\cite{DeAmicis:2026tus}), one obtains
\begin{align}\label{eq:allcontributions}
 G_{\ell m}(t-t',r_*,r_*')=&\notag -{1 \over 2\pi}\int_{\mathcal{C}_{\rm BC}}d\omega \, \tilde{G}_{\ell m}(\omega,r_*,r_*') e^{-i \omega (t-t')}\\&\  -{1 \over 2\pi}\int_{\mathcal{C}_{0}}d\omega\, \tilde{G}_{\ell m}(\omega,r_*,r_*') e^{-i \omega (t-t')}-i\sum_{\rm QNMs} \mathrm{Res}\big[ \tilde{G}_{\ell m}(\omega,r_*,r_*'), \omega_n \big]e^{-i \omega_n (t-t')}\,.
\end{align}

\subsubsection{A causal decomposition}
In the presence of a source, as defined above, the waveform can be written in general as
\begin{equation}
    \Psi_{\ell m}(t,r)=\int_{2M}^{\infty} dr'\int_{t_0}^{t}dt'\,G_{\ell m}(t-t',r_*,r_*') S_{\ell m}(r',t')\,.
\end{equation}
In this work, we consider the following decomposition of the response, isolating in particular the near--horizon contribution of the source from the contribution arising at larger values of $r'/M$. As a reference scale for this separation, we adopt the light ring location $r'=3M$, which lies close to the maximum of the effective potential barrier associated with the wave equation under consideration. Hence,
\begin{equation}
    \Psi_{\ell m}(t,r)=\int_{2M}^{3M} dr'\int_{t_0}^{t}dt'\,G_{\ell m}(t-t',r_*,r_*') S_{\ell m}(r',t')+\int_{3M}^{\infty} dr'\int_{t_0}^{t}dt'\,G_{\ell m}(t-t',r_*,r_*') S_{\ell m}(r',t')\,.
\end{equation}
This decomposition will allow us to clarify several structural features of the signal that are directly connected to its causal properties.

\subsubsection{Numerical code}
To test the analytical predictions derived in this work, we present numerical solutions of the Regge-Wheeler/Zerilli problem in Eq.~\eqref{eq:RWZ_equation}, for two different cases: an impulsive source as in Eq.~\eqref{eq:GF_equation}, to solve directly for the Green's function; a test-particle driving the perturbations, using the source in Eq.~\eqref{eq:testparticle_source}.
The solutions are computed using the \textsc{RWZHyp} code~\cite{Bernuzzi:2010ty,Bernuzzi:2011aj}; the code is characterized by a homogeneous grid in $r_*$ inside which the compact source is non-vanishing.
The grid is cut for a certain negative value of $r_*$, large in absolute value, so that the horizon is not included in the computational domain.
At large distances, the grid is attached to a hyperboloidal layer. 
The layer is parametrized by the retarded time $u$ and the compactified coordinate $\rho$, function of $r_*$, so that $\mathcal{I}^+$ is brought at a finite location denoted as $\rho_+$. It is possible to compute the signal at $\mathcal{I}^+$ without need for extrapolation.
The layer coordinates $(u,\rho)$ are related to the standard computational domain coordinates $(t,r_*)$ through
\begin{equation}
    u-\rho=t-r_* \, .
\end{equation}
The code uses double precision operations; we use the same resolution as in Ref.~\cite{DeAmicis:2024not}, to which we refer for an in-depth investigation of the code convergence.
When solving for the Green's function, we approximate the Dirac delta source in Eq.~\eqref{eq:GF_equation} through narrow Gaussians.
Analogously, we use a Gaussian to approximate the Dirac delta in the test-particle source in Eq.~\eqref{eq:testparticle_source}. 
This source is located along the test-particle trajectory, obtained by solving the Hamiltonian equations of motions driven by the analytical radiation-reaction effective forces derived in Refs.~\cite{Chiaramello:2020ehz,Albanesi:2021rby}.
We refer to Appendix~\ref{app:source} and Ref.~\cite{Nagar:2006xv} for their explicit expressions.
In this work, we will analyze two different planar trajectories of a test-particle with mass $\mu$: a radial infall from $r_0=50M$, with $\mu$-rescaled initial energy 
$E_0=1.00$; an eccentric inspiral with initial eccentricity~\footnote{We define the eccentricity through the location of the apastron and the periastron, $r_{\pm}$, as $e\equiv\frac{r_+-r_-}{r_++r_-}$} $e_0=0.5$, $\mu$-rescaled initial energy and $\mu$-rescaled angular momentum $E_0= 0.9587 $ and $p_{\varphi,0}=3.6502$, respectively.
Note that the eccentricity remains approximately constant throughout the inspiral evolution, such that at the separatrix crossing (after which no stable bound orbits exist and the system transitions to plunge) it holds $e_{\rm sep}=0.483$.

\section{Source outside the light ring}\label{sec:sourceinside}

\begin{figure}[ht]
    \centering    \includegraphics[width=0.92\linewidth]{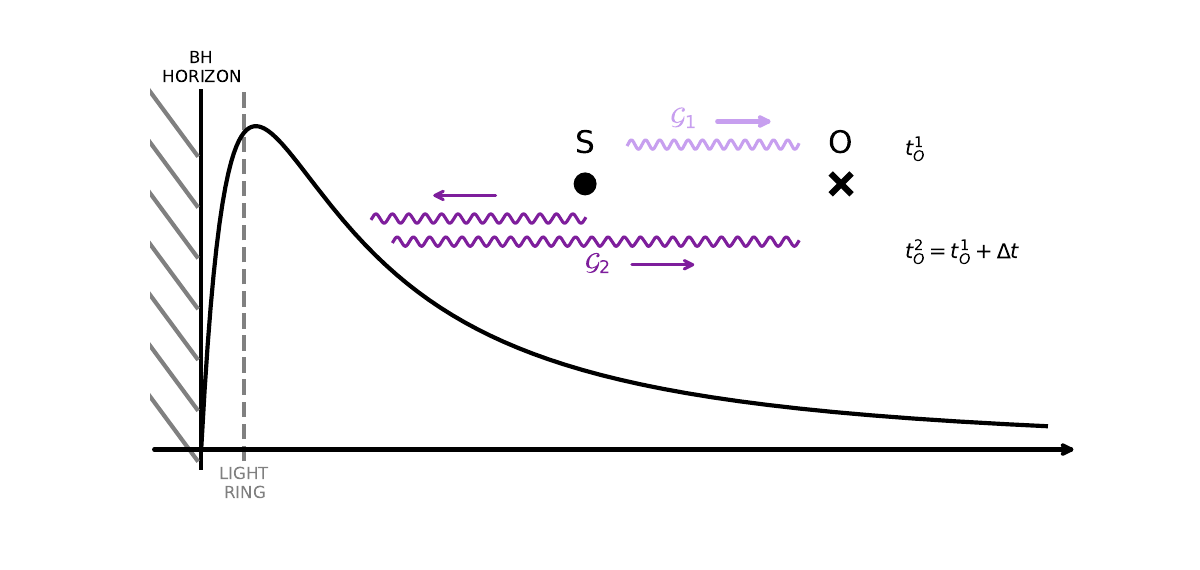}
    \caption{
    Schematic picture of the two causally distinct contributions $G^{(1)}_{\ell m}$ and $G^{(2)}_{\ell m}$ for a source $S$ supported outside the light ring ($r'>3M$) and an observer $O$ at future null infinity, $O\in\mathscr{I}^+$. The effective potential barrier peaked near the light ring acts as a scattering region. The direct channel, associated with $G^{(1)}_{\ell m}$, propagates outward from $S$ to $O$ without probing the barrier and therefore activates the prompt response at $t_0^{1}$. The scattered channel, associated with $G^{(2)}_{\ell m}$, initially propagates inward, interacts with the barrier, and is backscattered to infinity. It reaches $O$ only after a time delay $\Delta t=t_0^{2}-t_0^{1}=2r_*'-4M\log\!\big(f(r')\big)$ (see Eq.~\eqref{eq:delay} and Sec.~\ref{sec:greybody}). This causal delay underlies the fact that the ringdown contribution (quasinormal ringing plus late-time tails) is governed by $G^{(2)}_{\ell m}$ and is modulated by the reflection coefficient, directly related to the greybody factor.}
    \label{fig:causal_G1G2}
\end{figure}

In the region $r'>3M$, naturally one can divide the Green's function into two contributions. In particular, placing the observer at spatial infinity, we evaluate
Eq.~\eqref{eq:greenfunct} in the limit $r\to+\infty$.
Using the identity~\cite{Leaver:1986gd}
\begin{equation}\label{eq:divisiongreen} 
u^{\rm in}_{\ell m\omega}
= A^{\rm in}_{\ell m\omega}\, u^{\rm down}_{\ell m\omega}
+ A^{\rm out}_{\ell m\omega}\, u^{\rm up}_{\ell m\omega},
\end{equation}
and following Ref.~\cite{DeAmicis:2025xuh}, we obtain
\begin{equation}
\tilde{G}_{\ell m}(r_*,r'_*,\omega)
 = \frac{i\,e^{i\omega r_*}}{2\omega}\,
   u^{\rm down}_{\ell m\omega}(r'_*)
 + \frac{A^{\rm out}_{\ell m\omega}}{A^{\rm in}_{\ell m\omega}}
   \frac{i\,e^{i\omega r_*}}{2\omega}\,
   u^{\rm up}_{\ell m\omega}(r'_*)\,.
\end{equation}
We can then define (see Fig.~\ref{fig:causal_G1G2} for a schematic picture)
\begin{equation}\label{eq:G1G2}
\tilde{G}^{(1)}_{\ell m}(r_*,r'_*,\omega)
= \frac{i\,e^{i\omega r}}{2\omega}\,
  u^{\rm down}_{\ell m\omega}(r'_*)\,,
\qquad
\tilde{G}^{(2)}_{\ell m}(r_*,r'_*,\omega)
= \frac{A^{\rm out}_{\ell m\omega}}{A^{\rm in}_{\ell m\omega}}
  \frac{i\,e^{i\omega r}}{2\omega}\,
  u^{\rm up}_{\ell m\omega}(r'_*)\,.
\end{equation}

The two contributions $G^{(1)}_{\ell m}$ and $G^{(2)}_{\ell m}$ have a clear
physical interpretation as the direct response propagating to the observer and
the component generated by the interaction with the effective potential,
respectively. While Ref.~\cite{DeAmicis:2025xuh} showed that these two contributions are causally disconnected at the level of the QNM poles, and Ref.~\cite{DeAmicis:2026tus} established the same result for the prompt response, the structure of the corresponding late-time contribution has not yet been analyzed in detail. 

Although the decomposition in Eq.~\eqref{eq:divisiongreen} is formally valid everywhere, 
its physical interpretation requires some care. In particular, the identification of 
$G^{(1)}_{\ell m}$ as a direct component propagating to infinity and $G^{(2)}_{\ell m}$ 
as a contribution generated by scattering off the effective potential is meaningful 
only when the source lies outside the peak of the potential barrier. In this case, a 
portion of the radiation can propagate directly to infinity without interacting with 
the potential barrier, while the remaining part is produced by backscattering off the barrier. 
For the Schwarzschild potential this condition is approximately satisfied for 
$r' \gtrsim 3M$. Conversely, when the source lies inside the light ring ($r'<3M$), all 
radiation reaching infinity must first tunnel through the potential barrier, so that 
no component can propagate directly to infinity and the distinction between a direct 
and a scattered contribution loses its physical meaning.

\subsection{Late-time contribution}
In the following, we first compute the late-time tail of the full retarded Green's function for an observer at future null infinity, extending the small-frequency expansion beyond leading order and deriving the hierarchy of subleading corrections. We then introduce the causal decomposition $G_{\ell m}=G^{(1)}_{\ell m}+G^{(2)}_{\ell m}$ and show how to isolate the corresponding contributions to the late-time response. Finally, we demonstrate that the tails generated by $G^{(1)}_{\ell m}$ and $G^{(2)}_{\ell m}$ remain causally disconnected, reaching the observer with a relative time delay.

We present the results for the odd-parity sector and then generalize them to the even-parity sector.
\subsubsection{Late-time tails beyond leading order: odd sector}\label{sec:tails}
The late-time behavior of linearized perturbations is mainly controlled by the propagator's
non-analytic structure in the complex frequency plane. In particular, late-time
tails originate from the multivaluedness of the Green's function and are associated
with the branch cut extending along the negative imaginary-frequency axis~\cite{Leaver:1986gd}.
Denoting by $G^{\rm BC}_{\ell m}$ the branch-cut contribution (see Fig.~\ref{fig:GF_singular_structure}), one
can write
\begin{equation}\label{eq:GBC}
G^{\rm BC}_{\ell m}(t-t',r_*,r_*')
=
\frac{1}{2\pi}
\int_{\rm -i \infty}^{0} d\omega\,
\bigl[
\tilde{G}_{\ell m}(r_*,r'_*,\omega e^{2\pi i})
-
\tilde{G}_{\ell m}(r_*,r'_*,\omega)
\bigr]\,
e^{-i\omega(t-t')}\,,
\end{equation}
where the integration is performed along the branch cut on the negative imaginary
axis. 
The low-frequency behavior of the integrand in Eq.~\eqref{eq:GBC} gives the dominant contribution to the late-time behavior of the perturbations, i.e. for  $t-t' \gg M$.
For this reason, in the
following we perform a small-frequency expansion and show
how this regime gives rise to a hierarchy of late-time power-law decays, including
logarithmic corrections at subleading orders.

As anticipated, we consider the Regge-Wheeler equation describing axial perturbations,
\begin{equation}\label{eq:Regge-Wheeler}
r(r-2M)\,\tilde{\Psi}_{\ell m,rr} + 2M\tilde{\Psi}_{\ell m,r}
+ \left[\frac{\omega^{2} r^{3}}{r-2M} - \ell(\ell+1) + \frac{2M(s^{2}-1)}{r}\right]\tilde{\Psi}_{\ell m} = 0\,,
\end{equation}
where $\ell$ denotes the angular momentum number. Since our approach is valid for perturbations of any spin $s$, in the above equation, we keep $s$ generic (with $s=-2$ corresponding to axial gravitational perturbations).
Following Ref.~\cite{Leaver:1986gd,Leaver:1986vnb}, we perform the change of variables
\begin{equation}\label{eq:changevariables}
\tilde{\Psi}_{\ell m \omega}(r) = f(r)^{-2 i M \omega}\, h(z),
\end{equation}
with $f(r)=1-2M/r$ and $z=\omega r$, which transforms the Regge--Wheeler equation into the form
\begin{equation}\label{eq:coulomblike-Regge-Wheeler}
z\bigl(z-\omega \,r_h\bigr)\!\left[h_{zz} + \left(1 - \frac{2\eta}{z}\right) h_z\right]
+ C_1\,\omega\, h_z + \left(C_2 + \frac{C_3\,\omega}{z}\right) h = 0\,,
\end{equation}
with $\eta=-2 M \omega$, $r_h=2M$ and
\begin{equation}
C_1 = 2 M \bigl(1 - 4 i M \omega\bigr)\quad C_2 = -\,\ell - \ell^{2}+12 M^2 \omega^2 \quad C_3 = 2 M\left(-1 + s^{2} + 4 i M \omega + 4 M^{2}\omega^{2}\right)\,.    
\end{equation}

The resulting equation admits a solution expressible as a series expansion in Coulomb wave functions,
\begin{equation}\label{eq:coulombseries}
h(z) = \sum_{n=-\infty}^{+\infty} a_n\, u_{n+\nu}(\eta,z),
\end{equation}
where the coefficients $a_n$ and the renormalized angular momentum $\nu$ are solution of the recurrence relation specified in Appendix~\ref{app:recurrence}. The solutions introduced in the previous section can be written in terms of the
Coulomb functions $F_{\nu}(\eta,z)$ and $H^{(\pm)}_{\nu}(\eta,z)$.
The series involving $F_{\nu}(\eta,z)$ corresponds to $u^{\rm in}$, since it is
regular as $z\to 0$ and thus regular at the horizon, where $z_h=\omega r_h\to 0$
with the factor $f(r)^{-2iM\omega}$ from Eq.~\eqref{eq:changevariables} yielding a purely ingoing $\Psi_{\ell m}$ mode in this limit.
On the other hand, $H^{(+)}_{\nu}(\eta,z)$ ($H^{(-)}_{\nu}(\eta,z)$) corresponds to
$u^{\rm up}$ ($u^{\rm down}$), as it represents a right-moving (left-moving) plane
wave at spatial infinity.

For the up- and down-going solutions $u^{\rm up/down }_{\ell m\omega}$, one can write
\begin{equation}\label{eq:psiinf}
u^{\rm up/down }_{\ell m\omega}
= (2M\omega)^{\pm i\eta}\,e^{\pm i\phi_{\pm}}\,f(r)^{- 2Mi\omega}
\sum_{n=-\infty}^{+\infty}
a_{n}\,H^{(\pm)}_{n+\nu}(\eta,z)\,,
\end{equation}
where the phase $\phi_{\pm}$ is fixed so as to enforce the boundary condition
\eqref{eq:hom_sol2}, making use of the asymptotic form
\begin{equation}
    H^{(\pm)}_{n+\nu}(\eta,z\to \infty)\sim e^{\pm i \theta_{n+\nu}(\eta,z)}\,,
\end{equation}
with
\begin{equation}
    \theta_{n+\nu}(\eta,z)= z - \eta \log(2z) - (n+\nu){\pi \over 2} + \sigma_{n+\nu}(\eta)\,,
    \quad\quad\quad
    e^{i\sigma_{n+\nu}(\eta)}= \sqrt{\Gamma(n+\nu+1+i\eta) \over \Gamma(n+\nu+1-i\eta)}\,.
\end{equation}
The resulting expression for the phase reads
\begin{equation}\label{eq:phi}
\phi_{\pm}
= \pm i\,\ln\!\left[
\sum_{n=-\infty}^{+\infty}
a_{n}\,
\left(\frac{\Gamma(n+\nu+1 +i\eta)}{\Gamma(n+\nu+1 -i\eta)}\right)^{\!\mp1/2}
e^{\mp\,i (n+\nu)\pi/2}
\right].
\end{equation}

For $u^{\rm in}_{\ell m\omega}$ we write
\begin{equation}\label{eq:psirhmin}
u^{\rm in}_{\ell m\omega}(r_*) =N_{\ell m}(\omega)\, f(r)^{-2Mi\omega}\,\sum_{n=-\infty}^{+\infty}a_n\,F_{n+\nu}(\eta,z)\,,
\end{equation}
where the normalization $N_{\ell m}(\omega)$ can be fixed by imposing the boundary condition~\eqref{eq:hom_sol} at the horizon, although its
explicit form will not be required for our computation.

Since $u^{\rm in}_{\ell m\omega}(r)=A^{\rm out}_{\ell m \omega}u^{\rm up}_{\ell m\omega}(r)+A^{\rm in}_{\ell m \omega}u^{\rm down}_{\ell m\omega}(r)$, by comparing the two expressions we obtain
\begin{equation}
 A^{\rm out}_{\ell m \omega}={N_{\ell m}(\omega) \over 2i (2M \omega)^{i\eta} e^{i\phi_+}} \quad\quad  A^{\rm in}_{\ell m \omega}=-{N_{\ell m}(\omega) \over 2i (2M \omega)^{-i\eta} e^{-i\phi_-}}  \,. 
\end{equation}

To evaluate the integral along the branch cut on the negative imaginary-frequency
axis, we must determine how the solutions behave under the complex rotation
$\omega \to \omega e^{2\pi i}$.
One finds that $u^{\rm in}_{\ell m\omega}$ is single valued, whereas
$u^{\rm up/down}_{\ell m\omega}$ are multivalued.
This nontrivial monodromy originates from the Coulomb functions
$H^{(\pm)}_{n+\nu}(\eta,z)$.

In particular, the multivaluedness arises from the dependence on the variable $z$,
since the functions are single valued in both $n+\nu$ and $\eta$. To make this explicit, it is useful to express the Coulomb wave functions in terms of
confluent hypergeometric functions, namely the Kummer functions~\cite{DLMF}
Using this representation
\begin{equation}\label{eq:Hdecomp}
    H^{(\pm)}_{n+\nu}(\eta,z)=e^{\pm \theta_{n+\nu}(\eta,z)}(\mp 2iz)^{n+\nu+1\pm i\eta}U(n+\nu+1\pm i\eta,2n+\nu+2,\mp2iz),
\end{equation}
together with the analytic continuation formula
\begin{equation}
U\!\left(a,b, z e^{2\pi i m}\right)
=
\frac{2\pi i\, e^{-\pi i b m}\, }
{\Gamma(1+a-b)\, \Gamma(b)}\,
M(a,b,z)
+
e^{-2\pi i b m}\, U(a,b,z),
\end{equation}
and the relation
\begin{equation}\label{eq:rel1M}
\frac{1}{\Gamma(b)}\, M(a,b,z)
=
\frac{e^{ a \pi i}}{\Gamma(b-a)}\, U(a,b,z)
+
\frac{e^{ (b-a)\pi i}}{\Gamma(a)}\, e^{z}\,
U\!\left(b-a,\, b,\, e^{ \pi i} z\right),
\end{equation}
one finds that the up-going solution transforms as
\begin{equation}\label{eq:jump1}
    u^{\rm up}_{\ell m}(\omega e^{2\pi i},r)
    =
    u^{\rm up}_{\ell m}(\omega,r)
    -
    K(\omega)\,u^{\rm down}_{\ell m}(\omega,r),
\end{equation}
and analogously the down-going solution
\begin{equation}\label{eq:jump2}
    u^{\rm down}_{\ell m}(\omega e^{2\pi i},r)
    =
    u^{\rm down}_{\ell m}(\omega,r)
    -
    K(-\omega)\,u^{\rm up}_{\ell m}(\omega,r),
\end{equation}
where
\begin{equation}\label{eq:komega}
    K(\omega)
    =
    \left(e^{-2\pi (i\nu+\eta)}-1\right)
    \frac{A^{\rm in}_{\ell m \omega}}{A^{\rm out}_{\ell m \omega}}\,.
\end{equation}

In order to compute the jump of the full Green's function in
Eq.~\eqref{eq:greenfunct}, it is also necessary to determine how the coefficients
$A^{\rm in}_{\ell m \omega}$ and $A^{\rm out}_{\ell m \omega}$ transform across the branch cut.
This can be obtained by considering the relations
\begin{equation}
    -2i\omega A^{\rm in}_{\ell m}(\omega)=\mathcal{W}(u^{\rm up}_{\ell m},u^{\rm in}_{\ell m})\,,\quad  2i\omega A^{\rm out}_{\ell m}(\omega)=\mathcal{W}(u^{\rm down}_{\ell m},u^{\rm in}_{\ell m})\,,
\end{equation}
from which, using Eqs.~\eqref{eq:jump1} and~\eqref{eq:jump2}, it follows that
\begin{equation}
  A^{\rm in}_{\ell m}(\omega e^{2 \pi i})
  =
  A^{\rm in}_{\ell m}(\omega)
  +
  K(\omega)A^{\rm out}_{\ell m}(\omega)\,,
\end{equation}
and
\begin{equation}
    A^{\rm out}_{\ell m}(\omega e^{2 \pi i})
  =
  A^{\rm out}_{\ell m}(\omega)
  +
  K(-\omega)A^{\rm i n}_{\ell m}(\omega)\,.
\end{equation}

As a consequence, one readily finds
\begin{equation}
    \tilde{G}_{\ell m}(r_*,r'_*,\omega e^{2\pi i})
    -
    \tilde{G}_{\ell m}(r_*,r'_*,\omega)
    =
    \frac{i \left(1-e^{+2\pi (i\nu+\eta)}\right)}
    {2 M\omega A^{\rm out}_{\ell m}(\omega)A^{\rm in}_{\ell m}(\omega)}
    u^{\rm in}_{\ell m}(\omega,r)\,
    u^{\rm in}_{\ell m}(\omega,r')\,.
\end{equation}

Placing the observer at spatial infinity, the ingoing solution can be written as
\begin{equation}
    u^{\rm in}_{\ell m}(\omega,r)
    =
    A^{\rm out}_{\ell m}(\omega)e^{i\omega r_*}
    +
    A^{\rm in}_{\ell m}(\omega)e^{-i\omega r_*}\,.
\end{equation}

Upon performing the integration in Eq.~\eqref{eq:timegreenfunct}, one can show that
the first term generates late-time tails in the retarded time
$u=t-r_*$, while the second produces contributions in the advanced time
$v=t+r_*$.
In the regime $t,r_*\gg M$, the latter are always negligible and can therefore be
discarded.

Accordingly, we retain
\begin{align}
   \tilde{G}^{\rm BC}_{\ell m}(r_*,r'_*,\omega)
    =\tilde{G}_{\ell m}(r_*,r'_*,\omega e^{2\pi i})
    -
    \tilde{G}_{\ell m}(r_*,r'_*,\omega)
    =\notag\hspace{8cm}\\
    2\,e^{i \omega r_*} (2M \omega)^{-1-i\eta} e^{-i\phi_-}  \left(1-e^{+2\pi (i\nu+\eta)}\right)
    f(r')^{-2iM\omega}\sum_{n=-\infty}^{+\infty} a_n F_{n+\nu}(\eta,r'\omega) \,.
\end{align}

This expression is well suited for a small-frequency expansion, which allows one to
compute the late-time tails.
Before entering the details of this expansion, it is useful to consider the following
relation:
\begin{equation}
F_{n+\nu}(\eta,z)
=
C_{n+\nu}(\eta)\,
z^{n+\nu+1}\,
e^{-iz}\,
M\!\left(n+\nu+1-i\eta,\; 2n+2\nu+2,\; 2iz\right)\,,
\end{equation}
where
\begin{equation}
    C_{n+\nu}(\eta)
= 2^{n+\nu}\,e^{-\pi \eta/2}\,
\frac{\big|\Gamma(n+\nu+1+i\eta)\big|}{\Gamma(2n+2\nu+2)}\,.
\end{equation}
We can further observe that
\begin{equation}\label{eq:reltortoise}
    e^{-iz}\left(f(r)\right)^{-2iM\omega} \left(2M\omega\right)^{-i\eta} z^{i\eta}
    =
    e^{-i\omega r_*(r')}\,,
\end{equation}
which allows us to rearrange the branch-cut contribution as
\begin{align}\label{eq:BCgreentoexpand_I}
    &\tilde{G}^{\rm BC}_{\ell m}(r_*,r'_*,\omega)=\notag\\&\quad
    \,e^{i \omega \left(r_*-r'_*\right)} e^{-i\phi_-}
    {1-e^{+2\pi (i\nu+\eta)}\over M \omega\quad}
    \sum_{n=-\infty}^{+\infty}
    a_n C_{n+\nu}(\eta)
    (\omega r')^{n+\nu+1-i\eta}
    M\!\left(n+\nu+1-i\eta,\; 2n+2\nu+2,\; 2i\omega r'\right)\,.
\end{align}
Negative values of $n$ do not cause the series to diverge as $\omega \to 0$, since
$a_n \sim (M\omega)^{|n|}$, as discussed in Appendix~\ref{app:recurrence}.
The above expression can be readily expanded in powers of $\omega$, since for any
$b\neq -n$ with $n\in\mathbb{N}$ the confluent hypergeometric function admits the
series representation
\begin{equation}
    M(a,b,w)=\sum_{k=0}^\infty \frac{(a)_k}{(b)_k}\,\frac{w^k}{k!}\,,
\end{equation}
where $(q)_n=\frac{\Gamma(q+n)}{\Gamma(q)}$ denotes the Pochhammer symbol.
All the quantities entering Eq.~\eqref{eq:BCgreentoexpand_I} admit a Taylor expansion in $\omega$,
except for the factor $(\omega r')^{-i\eta}$, which yields the logarithmic series
\begin{equation}
(\omega r')^{-i\eta}
=\exp\!\bigl[-i\eta \log(\omega r')\bigr]
=1+\sum_{k=1}^{\infty}\frac{\bigl(2 i M\omega \,\log(\omega r')\bigr)^k}{k!}\,.
\end{equation}
As a consequence, the expansion of Eq.~\eqref{eq:BCgreentoexpand_I} takes the general form
\begin{equation}
\tilde{G}^{\rm BC}_{\ell m}(r_*,r'_*,\omega)
=\sum_{n=0}^{\infty}\sum_{k=0}^{n} g^{(n k)}_{\ell m}(r')\,(M\omega)^{n+\ell+1} \,\log(M\omega)^k \, .
\end{equation}
We can therefore compute the general form for the time-domain Green's function to be 
\begin{equation}
G^{\rm BC}_{\ell m}(t-t',r_*,r_*')
={1 \over 2\pi}\int_{-i\infty}^0 d\omega\,\tilde{G}^{\rm BC}_{\ell m}(r_*,r'_*,\omega)e^{-i \omega (t-t')}=\sum_{n=0}^{\infty}\sum_{k=0}^{n} g^{(n k)}_{\ell m}(r')\,I_{n+\ell,\,k}(u-u')
\end{equation}
where the explicit form of $I_{n+\ell,k}(u-u')$ is given in Appendix~\ref{app:timedependencetail} and yields to 
\begin{equation}\label{eq:greenintegral}
G^{\rm BC}_{\ell m}(u,u')
=\sum_{n=0}^{\infty}\sum_{k=0}^{n} \tilde{g}^{(n k),\,o}_{\ell m}(r')\,(u-u')^{-(\ell+n+1)}\log\left(u-u'\right)^k\,.
\end{equation}

Our derivation shows that, while the late-time signal measured
at $\mathscr{I}^{+}$ is governed at leading order by a pure power-law decay, the
presence of logarithmic corrections is unavoidable beyond leading order. In
fact, logarithmic terms arise systematically at each subleading order in the
expansion, reflecting the branch-cut structure of the Green's function in the
frequency domain. As a result, the late-time tail is characterized by a hierarchy
of power-law decays dressed by increasing powers of logarithmic terms.

The explicit expression of the coefficients for $\ell=-s=2$ perturbations is given in Appendix~\ref{app:coeffGreen}.
The code to compute perturbations for arbitrary $(l,s)$ is publicly available in the 
 \href{https://github.com/romeofelicerosato-prog/tails_higherorders}{\texttt{tails\_higherorders}} repository.

\subsubsection{Late-time tails in the even sector}
So far, we have focused on odd-parity perturbations. Nevertheless, all our results can be straightforwardly extended to the even-parity sector. To this end, we make use of the Chandrasekhar transformation~\cite{Chandrasekhar:1985kt}, which provides a direct mapping between the odd- and even-parity master functions. Explicitly, one has
\begin{equation}
\label{eq:Chandrasekhar_RW_to_Z}
\Psi^{e}_{\ell m\omega}(r)
=\mathcal{D}_+\Psi^{o}_{\ell m\omega}(r)=
\frac{1}{n(n+1)+3 i\omega M}
\left[
\left(
n(n+1)+\frac{9M^2 f(r)}{r(rn+3M)}
\right) \Psi^{o}_{\ell m\omega}(r)
+
3M f(r)\,\frac{d\Psi^{o}_{\ell m\omega}(r)}{dr}
\right]\,
\end{equation}
with \(n=\ell(\ell+1)/2\).

By inspecting the asymptotic behavior of Eq.~\eqref{eq:Chandrasekhar_RW_to_Z} at spatial infinity and imposing the boundary conditions given in Eq.~\eqref{eq:hom_sol2}, one readily finds that, with our normalization,
\begin{equation}
\label{eq:Ain_even}
A^{\rm in,\,(+)}_{\ell m}(\omega)
=
\frac{n(n+1)-3iM\omega}{n(n+1)+3iM\omega}
\,A^{\rm in,\,(-)}_{\ell m}(\omega),
\qquad
A^{\rm out,\,(+)}_{\ell m}(\omega)
=
A^{\rm out,\,(-)}_{\ell m}(\omega)\,.
\end{equation}

Equations~\eqref{eq:Chandrasekhar_RW_to_Z} and~\eqref{eq:jump1} then imply that the jump condition in the even-parity sector takes the particularly simple form
\begin{equation}
\label{eq:jump_even}
u^{\rm up,\,e}_{\ell m}(\omega e^{2\pi i},r)
=
u^{\rm up,\,e}_{\ell m}(\omega,r)
-
K(\omega)\,u^{\rm down,\,e}_{\ell m}(\omega,r),
\end{equation}
where \(K(\omega)\) is the same function defined in Eq.~\eqref{eq:komega}.

Making use of these relations, and closely following the derivation presented in the previous section, one finds that the branch-cut contribution to the Green's function in the even-parity sector is given by
\begin{align}
\tilde{G}^{\rm BC\,,e}_{\ell m}(r_*,r'_*,\omega)=
2\,e^{i \omega r_*} (2M \omega)^{-1-i\eta} e^{-i\phi_-}
\frac{e^{-2\pi (i\nu+\eta)}-1}
{e^{i \alpha }(e^{i\alpha}-1+e^{-2\pi (i\nu+\eta)})}
\, g(r',\omega)\,,
\end{align}
with
\begin{equation}
g(r',\omega)=
\mathcal{D}_+
\left(
f(r')^{-2Mi\omega}
\sum_{n=-\infty}^\infty a_n
F_{n+\nu}(\eta,r'\omega)
\right)\,.
\end{equation}

This expression can be expanded as in the odd-parity case. As a result, the even-parity branch-cut Green's function can again be written in the generic form
\begin{equation}
\label{eq:greenintegraleven}
G^{\rm BC\,,e}_{\ell m}(u,u')
=
\sum_{n=0}^{\infty}\sum_{k=0}^{n}
\tilde{g}^{(n k),\,e}_{\ell m}(r')\,
(u-u')^{-(\ell+n+1)}
\log\!\left(u-u'\right)^k\,.
\end{equation}

See Appendix~\ref{app:coeffGreen} for an explicit expression of the coefficients for $\ell=-s=2$ perturbations. The code to compute perturbations for arbitrary $(l,s)$ is publicly available in the 
\href{https://github.com/romeofelicerosato-prog/tails_higherorders}{\texttt{tails\_higherorders}} repository.

\subsubsection{Numerical validation: full Green's function and selected trajectories}
\label{sec:num_validation}

We now validate the analytic late-time expansion of the full retarded Green's function
against numerical computations at future null infinity. We consider axial perturbations, for which our analytic prediction is
given by Eq.~\eqref{eq:greenintegral}. The latter captures both the leading power-law decay
and the hierarchy of subleading corrections, including logarithmic terms, beyond
leading order.
The numerical waveforms are computed using the \textsc{RWZHyp} code~\cite{Bernuzzi:2010ty,Bernuzzi:2011aj}, as detailed in Sec.~\ref{sec:framework}.

A direct comparison with the raw numerical signal is limited by the oscillatory QNM content that dominates at intermediate times. To isolate the
tail, we therefore apply a rational filter that removes a prescribed number of QNMs, first developed by Ref.~\cite{Ma:2022wpv}, to which we refer for details on the filter implementation. 
This procedure yields a filtered signal in which the late-time branch-cut contribution is present already at relatively early retarded times.

\begin{figure}[h]
    \centering    \includegraphics[width=\linewidth]{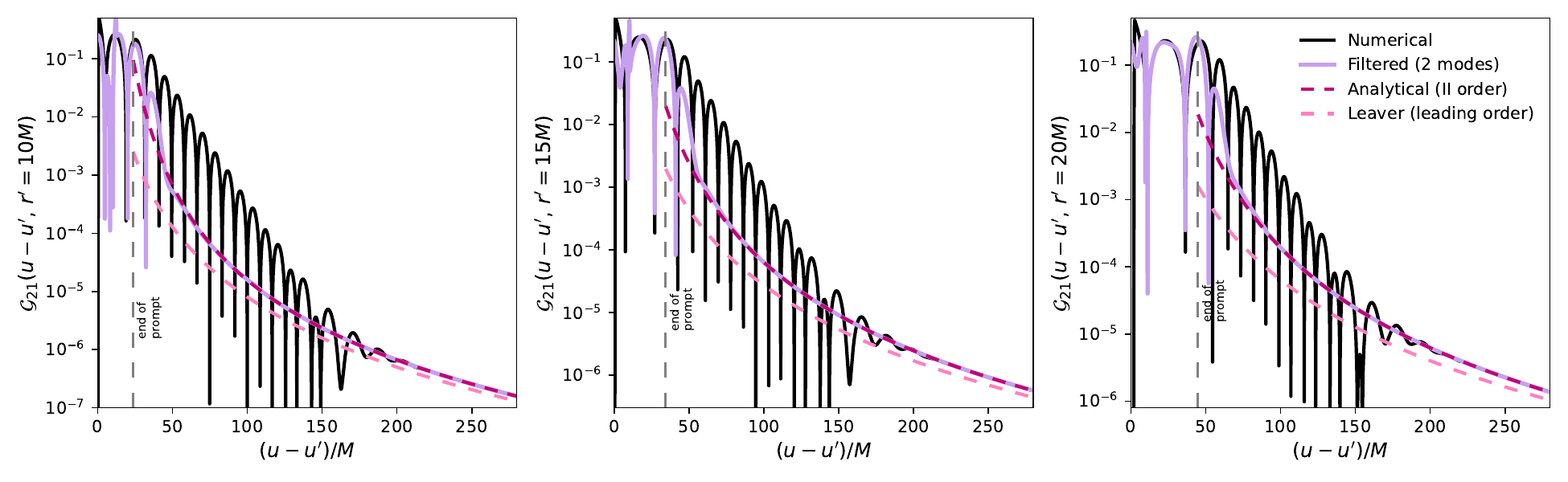}
    \caption{ Late-time response of the full Green's function at $\mathscr{I}^+$
    for several source locations $r'$ with $r'>3M$ for the $(\ell=2,m=1)$ axial mode.
    Black: raw numerical signal.
    Purple: rationally filtered signal after removing two QNM poles (fundamental mode and first overtone).
    Pink: Leaver's leading order result for the power-law tail.
    Magenta (dashed): analytic late-time expansion of the tail based on Eq.~\eqref{eq:greenintegral},
    truncated to leading order plus the first two subleading orders.}
    \label{fig:tail_fullGF_r0}
\end{figure}

Figure~\ref{fig:tail_fullGF_r0} shows representative results for the full Green's function for a set of source locations $r'$ in the region $r'>3M$. The filtered waveform exhibits excellent agreement with the analytic tail, not only in the asymptotic regime but also down to times that are comparable to the nominal end of the prompt/ringdown transition.
In particular, the beyond-leading-order expression reproduces the filtered decay significantly better than the leading-order Leaver's prediction alone, highlighting the quantitative relevance of the subleading corrections derived in Sec.~\ref{sec:tails}.
Here and in the plots below we include the leading order plus the first two subleading orders of the tail. 

From Fig.~\ref{fig:tail_fullGF_r0} two key points emerge. First, already about $25M$ after the end of the prompt response, subtracting only the fundamental mode and the first overtone leaves a residual signal dominated by the tail. Second, neglecting the subleading corrections derived here would lead to an underestimate of the theoretical tail amplitude in the Green's function by nearly one order of magnitude.

Having validated the analytic expression at the level of the full Green's function, we next consider specific source trajectories. For each trajectory, we compute the corresponding response through the convolution of the Green's function and the source, and apply the same rational filtering strategy to cleanly expose the tail.

We consider two representative trajectories: an eccentric infall with initial eccentricity $e_0=0.5$ and a radial infall starting from $r_0=50M$ (more details on the initial binary configurations can be found in Sec.~\ref{sec:framework}). The tail is known to be enhanced for larger eccentricities, and in particular for radial infall ($e=1$) from more distant initial positions~\cite{DeAmicis:2024not}. 

\begin{figure}[h]
    \centering
    \includegraphics[width=\linewidth]{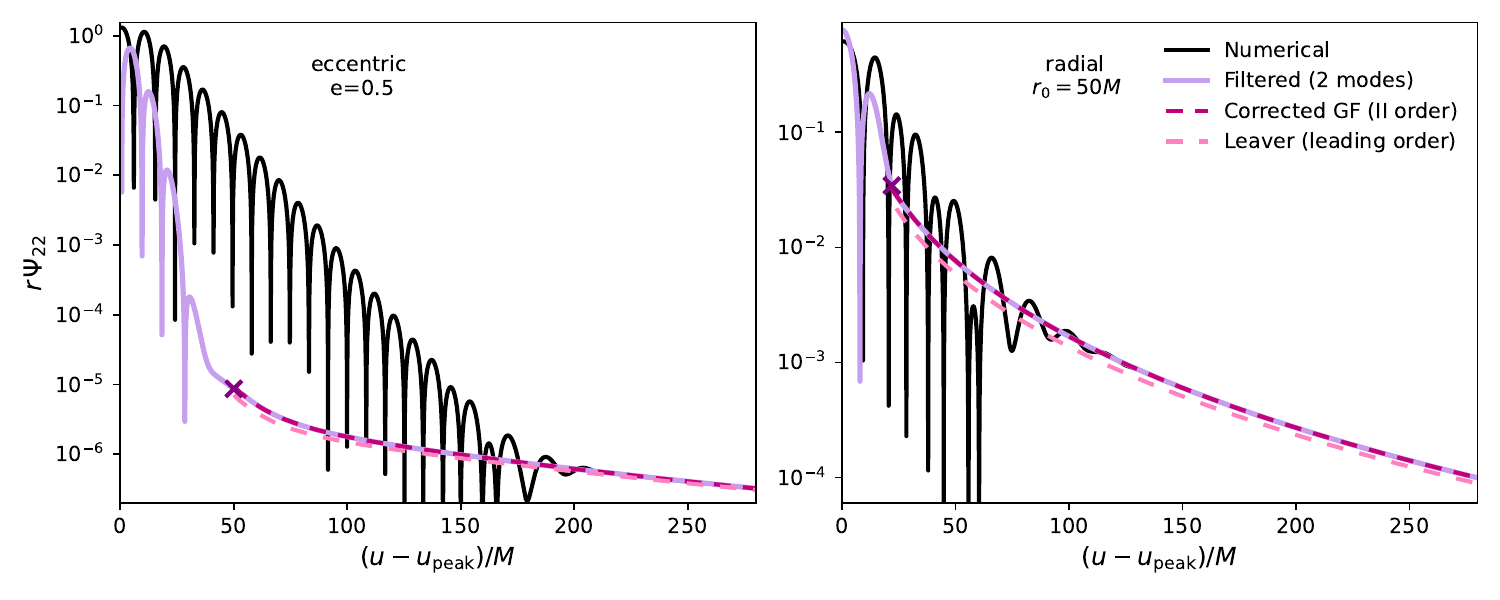}
    \caption{Late-time response along selected source trajectories for the (22) mode. We consider one eccentric trajectory ($e_0=0.5$) and a radial infall.
    Black: numerical waveform. Purple: rationally filtered signal, after the removal of 2 QNMs (fundamental mode and first overtone). Magenta (dashed):
    analytic tail from Eq.~\eqref{eq:greenintegral} (leading order plus the first two
    subleading orders). The analytic prediction accurately tracks the filtered tail
    down to times that overlap with the ringdown-dominated regime.}
    \label{fig:tail_trajectories}
\end{figure}

As shown in Fig.~\ref{fig:tail_trajectories}, the agreement with the analytic prediction remains robust for the considered trajectories. The impact of subleading contributions is smaller than in the full Green's function case. 
The reason lies in a difference between timescales. The enhanced tail observed at intermediate and late times is generated when the source (be it test particle or Gaussian packet) is at large distances. So in the test-particle case, for eccentric orbits, the tail is mainly emitted at the last apastron. In the Gaussian packet case, the tail is emitted at the location of the impulsive source. The time interval between the ``loud'' tail emission and ringdown is longer in the test-particle case, than in the Gaussian packet one. Hence, by the time the tail emitted at the last apastron is observed, faster decaying corrections have considerably decayed and have a smaller influence on the full signal. 
Nevertheless, in the test-particle case, including the first two corrective orders leads to noticeably better agreement with the numerical data. Not including them yields an underestimation of the tail magnitude under the ringdown. This effect is stronger for the radial infall case (since it is characterized by faster dynamics).
The prediction in Eq.~\eqref{eq:greenintegral} improves our modeling capabilities by providing a faithful description of the tail already at intermediate times. This allows us to probe a regime in which the signal amplitude is significantly larger than in the late-time domain, where the tail eventually dominates the strain.
In the intermediate-eccentricity case, $e_0=0.5$, the higher-order corrected tail overtakes the first two QNMs and their mirror modes at $\sim 50M$ after the light-ring crossing. At that time its amplitude is roughly one order of magnitude larger than when the tail becomes clearly visible in the waveform.
In the radial infall case, as expected, the tail is more strongly excited. It dominates over the first two QNMs and their mirror modes earlier, around $\sim 20M$ after the light-ring crossing. At that time its higher-order corrected amplitude is suppressed by only about two orders of magnitude relative to the peak of the full strain.

\subsubsection{Causal separation of the late-time tails}

At the beginning of this section, we showed that the total Green's function of the
problem can be decomposed into two contributions, $G^{(1)}_{\ell m}$ and $G^{(2)}_{\ell m}$, see Eq.~\eqref{eq:G1G2}. One may then ask whether the late-time tail associated
with each contribution can be computed separately. The answer is affirmative.
The key result of this section is that the two tails are causally disconnected:
the tail generated by $G^{(2)}_{\ell m}$ reaches the observer with a delay with respect to the tail associated with $G^{(1)}_{\ell m}$.

To demonstrate this property, we consider the expression of
$u^{\rm in}_{\ell m}(\omega,r')$ in terms of confluent hypergeometric functions,
\begin{equation}\label{eq:psirhminhyper}
u^{\rm in}_{\ell m\omega}(r_*) =N_{\ell m}(\omega)\, f(r)^{-2Mi\omega}\,\sum_{n=-\infty}^{+\infty}a_n\,C_{n+\nu}(\eta)\,
z^{n+\nu+1}\,
e^{-iz}\,
M\!\left(n+\nu+1-i\eta,\; 2n+2\nu+2,\; 2iz\right)\,.
\end{equation}
By using Eq.~\eqref{eq:rel1M}, one finds
\begin{align}\label{eq:MasdoubleU}
&e^{-iz}\,
M\!\left(n+\nu+1-i\eta,\; 2n+2\nu+2,\; 2iz\right)=
\left(e^{-iz-\pi\eta-i\pi(n+\nu+1)} {\Gamma(2n+2\nu+2)\,
 \over \Gamma(n+\nu+1+i\eta)}U\!\left(n+\nu+1-i\eta,\; 2n+2\nu+2,\; 2iz\right)\right.\notag\\&\hspace{4cm} +\left.e^{iz+\pi\eta+i\pi(n+\nu+1)}  {\Gamma(2n+2\nu+2)\,
 \over \Gamma(n+\nu+1-i\eta)}U\!\left(n+\nu+1+i\eta,\; 2n+2\nu+2,\; -2iz\right)\right)\,.
\end{align}
By further employing Eqs.~\eqref{eq:psiinf} and~\eqref{eq:Hdecomp}, it follows
straightforwardly that the first term in the above expression corresponds to
$u^{\rm down}_{\ell m}(\omega,r')$, while the second term corresponds to
$u^{\rm up}_{\ell m}(\omega,r')$. According to Eq.~\eqref{eq:G1G2}, these two
contributions therefore build $G^{(1)}_{\ell m}$ and $G^{(2)}_{\ell m}$, respectively
\begin{align}\label{eq:BCG1}
    &\tilde{G}^{(1/2),\,\rm BC}_{\ell m}(r_*,r'_*,\omega)=\notag\,2\,e^{i \omega r_*} (2M \omega)^{-1-i\eta} e^{-i\phi_-} \left(1-e^{+2\pi (i\nu+\eta)}\right)
    f(r')^{-2iM\omega}\times\\&\quad\quad
     \sum_{n=-\infty}^{+\infty} a_n C_{n+\nu}(\eta) z^{n+\nu+1} e^{\mp iz}  e^{\mp i\pi(n+\nu+1-i\eta)}{\Gamma(2n+2\nu+2)\,
 \over \Gamma(n+\nu+1\pm i\eta)}U\!\left(n+\nu+1\mp i\eta,\; 2n+2\nu+2,\; \pm2iz\right) \,.
\end{align}

At first sight, the representation of the Green's function modes in terms of the
confluent hypergeometric function $U(a,b,w)$ may appear problematic, since $U$
has poles in the low-frequency limit. However, this does not signal any
physical divergence. Indeed, these poles are such that they must cancel out
between the two contributions $G^{(1)}_{\ell m}$ and $G^{(2)}_{\ell m}$, since their sum is required
to reconstruct the confluent hypergeometric function $M(a,b,w)$, which is
regular at low frequencies.

This cancellation can be made explicit by using the identity
\begin{equation}\label{eq:UinTermsOfM}
U(a,b,w)
=
\frac{\Gamma(1-b)}{\Gamma(a-b+1)}\,M(a,b,w)
+
\frac{\Gamma(b-1)}{\Gamma(a)}\,w^{1-b}\,
M(a-b+1,2-b,w)\,,
\end{equation}
from which the pole structure of $U$ is entirely encoded in the second term.
However, these terms cancel exactly between $G^{(1)}$ and $G^{(2)}$
upon using the identity $M(a,b,w)=e^{w}\,M(b-a,b,-w)$.

As a consequence, the pole structure of $U$ does not contribute to the late-time
signal, but only to the early-time response~\cite{DeAmicis:2026tus}. The remaining terms that control the late-time behavior are therefore
given by
\begin{align}\label{eq:BCG1bis}
    & \tilde{G}^{(1/2),\,\rm BC}_{\ell m}(r_*,r'_*,\omega)=\notag\,2\,e^{i \omega r_*} (2M \omega)^{-1-i\eta} e^{-i\phi_-} \left(1-e^{+2\pi (i\nu+\eta)}\right)
    f(r')^{-2iM\omega}\times\\&\quad\quad
     \sum_{n=-\infty}^{+\infty} a_n C_{n+\nu}(\eta) z^{n+\nu+1} e^{-i\pi(n+\nu+1)} e^{\mp (iz+\pi\eta)} \sigma_\pm(\omega)M\!\left(n+\nu+1\mp i\eta,\; 2n+2\nu+2,\; \pm2iz\right) \,,
\end{align}
with
\begin{equation}
    \sigma_\pm (\omega)=\frac{\sin(n+\nu+1\pm i\eta)}{\sin(2n+2\nu+2)}\,.
\end{equation}

Note that, according to Appendix~\ref{app:recurrence}, the renormalized angular
momentum admits the low-frequency expansion
$\nu=\ell+\sum_j \nu_j (M\omega)^2$. Together with $\eta=-2M\omega$, this implies
that $\sigma_\pm$ develops a simple pole proportional to
$(-1)^{n+\ell+1}(\pm i\eta/2\nu_2)$ in the low-frequency limit.
However, this pole does not lead to any physical divergence. Indeed, when
retaining only the leading-order contribution of the above expression, the pole
cancels exactly, and can therefore be consistently discarded. In what follows, we implicitly work with the regularized quantities
$\sigma^{\rm REG}_\pm(\omega)$, obtained from $\sigma_\pm(\omega)$ by discarding
the spurious low-frequency pole described above.

By using Eq.~\eqref{eq:reltortoise} into Eq.~\eqref{eq:BCG1bis}, we obtain
\begin{align}\label{eq:BCgreentoexpand}
    &\tilde{G}^{(1),\,\rm BC}_{\ell m}(r_*,r'_*,\omega)=e^{i \omega \left(r_*-r'_*\right)}\notag e^{-i\phi_-}
    {1-e^{+2\pi (i\nu+\eta)}\over M \omega\quad}\times\\&\quad\quad\quad
    \, 
    \sum_{n=-\infty}^{+\infty}
    a_n C_{n+\nu}(\eta)
    (\omega r')^{n+\nu+1-i\eta} e^{-i\pi(n+\nu+1-i\eta)}\sigma^{\rm REG}_{+}(\omega)
    M\!\left(n+\nu+1-i\eta,\; 2n+2\nu+2,\; 2i\omega r'\right)\,,
\end{align}
and 
\begin{align}\label{eq:BCgreentoexpand}
    &\tilde{G}^{(2),\,\rm BC}_{\ell m}(r_*,r'_*,\omega)=e^{i \omega \left(r_*+r'_*\right)}e^{-4iM\omega\log\left(f(r')\right)}2e^{-i\phi_-}
    {1-e^{+2\pi (i\nu+\eta)}\over (2M \omega)^{1+2i\eta}\quad}\times\notag\\&\quad\quad\quad
    \, 
    \sum_{n=-\infty}^{+\infty}
    a_n C_{n+\nu}(\eta)
    (\omega r')^{n+\nu+1+i\eta} e^{i\pi(n+\nu+1-i\eta)}\sigma^{\rm REG}_{-}(\omega)
    M\!\left(n+\nu+1+i\eta,\; 2n+2\nu+2,\; -2i\omega r'\right)\,.
\end{align}

We observe that the two components of the Green's function differ by an overall
frequency-dependent phase,
$e^{i\omega \left(2r_*(r')-4M\log f(r')\right)}$.
When performing the inverse Fourier transform to the time domain, such a phase
translates into a shift of the retarded time argument, and therefore into a
delay of the corresponding late-time tail.

Proceeding as in the previous section, one finds that the branch-cut
contributions takes the form

\begin{equation}\label{eq:greenintegralG1}
G^{(1),\,\rm BC}_{\ell m}(u,u')
=\sum_{n=0}^{\infty}\sum_{k=0}^{n} \tilde{g}^{(1),(n k)}_{\ell m}(r')\,(u-u')^{-(\ell+n+1)}\log\left(u-u'\right)^k\,,
\end{equation}
and 
\begin{align}\label{eq:greenintegralG2}
G&^{(2),\,\rm BC}_{\ell m}(u,u')
=\notag\\&\quad\sum_{n=0}^{\infty}\sum_{k=0}^{n} \tilde{g}^{(2),(n k)}_{\ell m}(r')\,\left(u-u'+2r_*(r')-4M\log\left(f(r')\right)\right)^{-(\ell+n+1)}\log\left(u-u'+2r_*(r')+4M\log\left(f(r')\right)\right)^k\,,
\end{align}
respectively. Equation~\eqref{eq:greenintegralG2} shows that the late-time
response associated with $G^{(2)}_{\ell m}$ is shifted by an amount
\begin{align} \label{eq:delay}
    {\Delta t} =2r_*(r')-4M\log f(r')
\end{align}
with respect to the contribution arising from $G^{(1)}_{\ell m}$.
This shift has a clear causal interpretation. The late-time tails generated by
$G^{(1)}_{\ell m}$ and $G^{(2)}_{\ell m}$ are temporally separated, with the signal associated with
$G^{(2)}_{\ell m}$ reaching the observer only after an additional delay. As a result, the
two contributions are causally disconnected: at sufficiently late retarded
times, the tail of $G^{(1)}_{\ell m}$ has already decayed when the tail of $G^{(2)}_{\ell m}$
starts contributing.

Note that Ref.~\cite{Su:2026fvj} discusses the presence of distinct branch cuts associated with $G^{(1)}_{\ell m}$ and $G^{(2)}_{\ell m}$.
The approach adopted here is different but fully consistent: we retain the complete expression of the retarded Green's function when computing the discontinuity across the branch cut. In this framework, only a single branch cut along the negative imaginary-frequency axis appears.
The separation into the individual contributions of $G^{(1)}_{\ell m}$ and $G^{(2)}_{\ell m}$ is performed only at a later stage, once the total result has been obtained.

This procedure is also consistent with the full retarded Green's function expression for tails in Eq.~\eqref{eq:greenintegral}. Indeed, in the limit $u-u'\gg M$, we have
\begin{equation}
\left(u-u'+{\Delta t}\right)^{-(\ell+n+1)}
\log^{k}\!\left(u-u'+{\Delta t}\right)
=
\sum_{j\ge 0}
\sum_{p\le k}
\hat c^{(n,k)}_{j p}({\Delta t})\,
(u-u')^{-(\ell+n+1+j)}
\log^{p}(u-u') \,,
\end{equation}
which explicitly reproduces the same hierarchy of inverse powers of $(u-u')$ as in the expansion of the full Green's function.
\subsection{Frequency-Domain Causality and Greybody Factors}
\label{sec:greybody}
In this section we provide a unified interpretation of the causal separation between $G^{(1)}_{\ell m}$ and $G^{(2)}_{\ell m}$ for sources with support in $r'>3M$. In Sec.~\ref{sec:framework} we discussed how the complete time-domain response can be decomposed into three contributions: the QNM poles, the pole at $\omega=0$, and the branch-cut integral. 

With regard to causality and to the decomposition of the total Green's function into $G^{(1)}_{\ell m}$ and $G^{(2)}_{\ell m}$, we summarize here the key results that allow for a transparent interpretation of the Green's function in the frequency domain:
\begin{itemize}
    \item Ref.~\cite{DeAmicis:2025xuh} shows that the QNM contribution arises exclusively from $G^{(2)}_{\ell m}$. This component is therefore responsible for the quasinormal ringing, which is activated only after a time delay ${\Delta t}=2r_*'-4M\log(f(r'))$ with respect to the onset of the signal.
    \item Ref.~\cite{DeAmicis:2026tus} shows that the pole at $\omega=0$ accounts for the prompt response. 
    In particular, $G^{(1)}_{\ell m}$ carries the prompt contribution, which, after the time delay discussed above, is cancelled by the pole component of $G^{(2)}_{\ell m}$.
    \item In the previous sections we have completed this picture. We have shown that both $G^{(1)}_{\ell m}$ and $G^{(2)}_{\ell m}$ possess a late-time tail, since in both cases radiation backscatters off the effective potential. Furthermore, the tail of $G^{(2)}_{\ell m}$ appears only after a time delay of ${\Delta t}=2r_*'-4M\log(f(r'))$ with respect to that of $G^{(1)}_{\ell m}$. Consequently, the observable late-time signal is ultimately governed by $G^{(2)}_{\ell m}$, since the tail generated by $G^{(1)}_{\ell m}$ has already decayed when the tail of $G^{(2)}_{\ell m}$ becomes relevant.
\end{itemize}

Since we have verified this causal disconnection for all three components entering Eq.~\eqref{eq:allcontributions}, we conclude that such a causal separation holds at the level of the Green's function itself. In particular:
\begin{itemize}
    \item The Green's function is first activated at $t=t_0$ through the prompt response, entirely generated by $G^{(1)}_{\ell m}$. The signal is therefore governed by $G^{(1)}_{\ell m}$ for $t_0 < t < t_0 + 2r_*' -4M\log(f(r'))$.
    \item For $t > t_0 + 2r_*' - 4M\log(f(r'))$, during the ringdown stage, the signal is completely determined by $G^{(2)}_{\ell m}$.
\end{itemize}

This causal structure has a profound connection with greybody factors. Indeed, $G^{(2)}_{\ell m}$ can be written as
\begin{equation}
  \tilde{G}^{(2)}_{\ell m}(r_*,r'_*,\omega)= 
  \frac{A^{\rm out}_{\ell m\omega}}{A^{\rm in}_{\ell m\omega}}
  \frac{i\,e^{i\omega x}}{2\omega}\,
  u^{\rm up}_{\ell m\omega}(r'_*)
  \equiv  
  R_{\ell m\omega}\frac{i\,e^{i\omega x}}{2\omega}\,
  u^{\rm up}_{\ell m\omega}(r'_*)\,,
\end{equation}
so that it is entirely controlled by the reflection coefficient $R_{\ell m\omega}$, which is directly related to the greybody factor of the spacetime through Eq.~\eqref{eq:greybody}.

Consequently, in the frequency domain the portion of the signal associated with the ringdown (understood here as quasinormal ringing plus late-time tails) takes the form
\begin{equation}\label{eq:greybodyRD}
    \Psi_{\ell m\omega}(r)=
    R_{\ell m\omega}\frac{i\,e^{i\omega r_*}}{2\omega}
    \int dr' \,u^{\rm up}_{\ell m\omega}(r'_*)\, S_{\ell m\omega}(r')
    \equiv
    e^{i\omega r_*}\frac{R_{\ell m\omega}}{\omega}\,\rho_{\ell m}(\omega)\,,
\end{equation}
where $\rho_{\ell m}(\omega)$ depends only on the frequency and encodes the details of the source. This result provides a direct theoretical underpinning for recent works that phenomenologically model the frequency-domain signal as a greybody-modulated response, in which the reflectivity is imprinted in the amplitude and multiplied by a smooth frequency-dependent function~\cite{Oshita:2023cjz,Rosato:2024arw,Oshita:2024fzf,Rosato:2025byu,Rosato:2025ulx}.

This decomposition has a particularly transparent interpretation when viewed through a simple toy model. Consider a one-dimensional scattering problem governed by the inhomogeneous wave equation
\begin{equation}\label{eq:scattering1D}
\left[\frac{d^2}{dx^2} + \omega^2 - V(x)\right]\Psi(\omega,x)
= S(\omega,x)\,.
\end{equation}
Assume that $V(x)$ has support only for $x<0$, while the source $S(\omega,x)$ is localized to the right of this region. The response can then be
written as
\begin{equation}
\Psi(\omega,x)=\int_{-\infty}^{\infty} G(\omega;x,x')\,S(\omega,x')\,dx'\,,
\end{equation}
where $G(\omega;x,x')$ is the outgoing Green's function solving
Eq.~\eqref{eq:scattering1D}.  

Since $V(x)$ vanishes for $x>0$, the Green's function coincides with the
free propagator in this region, and its most general outgoing form is a
superposition of plane waves. Standard scattering theory then yields
\begin{equation}\label{eq:Glocalized}
G(\omega;x,x')=\frac{i}{2\omega}
\left[ e^{\,i\omega|x-x'|}
     + R(\omega)\,e^{\,i\omega(x+x')}\right],
\end{equation}
where $R(\omega)=A^{\rm out}(\omega)/A^{\rm in}(\omega)$ is the
frequency-domain reflection coefficient. For an observer at
$x\to+\infty$, this expression reduces to
\begin{equation}\label{eq:Glocalizedbis}
G(\omega;x,x')=\frac{i}{2\omega}
\left[ e^{\,i\omega(x-x')}
     + R(\omega)\,e^{\,i\omega(x+x')}\right].
\end{equation}

The physical interpretation is immediate. The first term,
$(i/2\omega)\,e^{\,i\omega(x-x')}$, represents the portion of the signal
generated by the right-moving component of the emitted signal. It propagates
freely from $x'$ to $x$ without interacting with the potential and thus
reaches the observer first, forming the direct part of the response.
The second term encodes the initially left-moving component of the signal, which
travels toward $x<0$, interacts with the potential $V(x)$, and is then
reflected back toward positive $x$. Its contribution at $x\to+\infty$
is therefore delayed, being proportional to
$(i/2\omega)R(\omega)e^{\,i\omega(x+x')}$.

This two-part structure is generic for localized scattering
problems and provides a simple physical picture underlying the
frequency-domain causal decomposition discussed above. We illustrate the black-hole case in Fig.~\ref{fig:causal_G1G2}. Consider an observer $O$ located at future null infinity ($\mathscr{I}^+$) and a source $S$ with support outside the light-ring, at a radius $r'>3M$. The peak of the effective potential, centered around the light-ring, plays the role of the localized scattering region of the toy model. The total response naturally splits into two distinct propagation channels. The component $G^{(1)}_{\ell m}$ of the Green's function propagates radiation directly outward from $S$ to $O$ without probing the potential barrier. Therefore it reaches $\mathscr{I}^+$ first, activating the prompt signal at time $t=t_0^{1}$. 

By contrast, $G^{(2)}_{\ell m}$ describes radiation that initially propagates inward, encounters the barrier near the light-ring, and is backscattered toward infinity. Its contribution at $\mathscr{I}^+$ is activated only at a later time $t=t_0^{2}=t_0^{1}+ {\Delta t}$. For sources with support in $r'>3M$, the delay is given by Eq.~\eqref{eq:delay}, which makes the causal separation between $G^{(1)}_{\ell m}$ and $G^{(2)}_{\ell m}$ explicit in the time domain. 

This provides a simple geometric interpretation of why the greybody factor (encoded in the reflection coefficient multiplying $G^{(2)}$) governs precisely the causally delayed portion of the observable signal.

\section{Source inside the light ring}\label{sec:sourceoutside}
In the region inside the light ring, namely $r'<3M$, the previous decomposition of the Green's function can still be performed formally, but it loses its clear physical interpretation.
From a physical perspective, this reflects the fact that the notion of reflection off the potential barrier ceases to be well defined. As a result, the separation into a direct and a reflected contribution no longer carries an unambiguous dynamical meaning, as we discuss in detail in Sec.~\ref{sec:gfsinsidelightring}.

It is therefore natural to consider the full Green's function without attempting any further subdivision. In the region $2M<r'<3M$, the solution can be expanded as a series in $r'/M-2$, which converges throughout the domain of interest. By imposing $u^{\rm in}_{\ell m \omega}(r)=e^{-i\omega r_*}h(r)$ and introducing the variable $x=r/2M -1$, the Regge--Wheeler equation can be recast as
\begin{equation}\label{eq:nearhor}
    -x (x+1)^2 h''(x)+i (x+1)\left(4 (x+1)^2 \omega +i\right)\,h'(x)+\left(l (l+1) (x+1)-3\right)\,h(x)=0\,.
\end{equation}

If we express $h(x)=\sum_{n=0}^\infty h_n x^n$ 
and substitute this expansion into the differential equation above, we obtain a recurrence relation for the coefficients $h_n$ for $n\geq 2$:
\begin{equation}\label{eq:recurrenceHor}
h_{n+1}
=
\frac{
\Bigl[-2n^2+n(1+12 iM\omega)+l(l+1)-3\Bigr]\,h_n
+\Bigl[l(l+1)+(n-1)(12 iM\omega-n+2)\Bigr]\,h_{n-1}
+4 iM\omega\,(n-2)\,h_{n-2}
}{
(n+1)\bigl(n+1-4 iM\omega \bigr)
}.
\end{equation}

The initial relations are given by
\begin{equation}
    h_1= \frac{i h_0 \left(l^2+l-3\right)}{4 M\omega +i}\,,\quad \quad 
    h_2= \frac{i \left((h_0+h_1) l (l+1)+(12 i M\omega -4)h_1 \right)}{8 M\omega +4 i}\,.
\end{equation}

The condition of a purely ingoing wave at the horizon is already built into this formalism through the ansatz for $u^{\rm in}_{\ell m \omega}$. The overall normalization is fixed by choosing $h_0$, which we set to unity. Taking the limit $n \to \infty$, one finds that $a_n/a_{n-1} \to (-1)^n$, so that the radius of convergence of the series is $1$. Therefore, the series converges for $r/M \in [2,3]$, i.e. up to the light ring.

By placing the observer at $\mathscr{I}^+$
, the total Green's function reads
\begin{equation}
\label{eq:greenHor}
\tilde{G}_{\ell m}(r_*,r'_*,\omega)
=
\frac{i\, e^{i\omega (r_* - r_*')}}{2\omega\, A^{\rm in}_{\ell m\omega}}\,
h\!\left(\frac{r'}{2M}-1\right)\,.
\end{equation}
Note that in this case there is no pole at $\omega=0$, since in the low-frequency approximation $A^{\rm in}_{\ell m\omega} \sim C_\ell\,\omega^{-\ell-1}$~\cite{Rosato:2025rtr}. Consequently, the Green's function is regular at $\omega=0$ in this representation, which is consistent with the statement that the prompt responses of $G^{(1)}_{\ell m}$ and $G^{(2)}_{\ell m}$ cancel each other out exactly.

\subsection{Branch cut contribution}
Here we derive the behavior of the Green's function along the branch cut. The only multi-valued quantity in Eq.~\eqref{eq:greenHor} is $A^{\rm in}_{\ell m\omega}$, whose jump across the branch cut has been derived in Sec.~\ref{sec:tails}. The computations involving $A^{\rm in}_{\ell m\omega}$ and its jump across the branch cut never assumed the source to be located in a specific region of spacetime, as this quantity (alongside with $A^{\rm in}_{\ell m\omega}$) is an invariant of the problem, being related to the Wronskian. Only exact solutions valid for all $r'$ were involved, hence they can be used also in this context. $h(x)$ is clearly a single-valued function of the frequency, as clear from the recurrence relation Eq.~\eqref{eq:recurrenceHor}.

By applying the results of the previous section, one finds
\begin{equation}
\tilde{G}^{\rm BC}_{\ell m}(r_*,r'_*,\omega)
=
\frac{i\, e^{i\omega (r_* - r_*')}K(\omega)A^{\rm out}_{\ell m\omega}}{2\omega\, A^{\rm in}_{\ell m\omega}(A^{\rm in}_{\ell m\omega}-K(\omega)A^{\rm out}_{\ell m\omega})}\,
h\!\left(\frac{r}{2M}-1\right)\,.
\end{equation}
We now focus on the leading-order behavior in $\omega$. Using the results of the previous section, one obtains
\begin{equation}
\tilde{G}^{\rm BC}_{\ell m}(r_*,r'_*,\omega)=2\pi h\!\left({r' \over 2M}-1\right)\Bigg|_{\omega=0}{e^{i\omega (r_*-r_*')} (2Mi\omega)^{\ell+1}\over (2\ell+1)!!}\,.
\end{equation}
Consequently, in the time domain one finds
\begin{equation}
G^{\rm BC}_{\ell m}(r,r',t-t')
=
h\!\left({r' \over 2M}-1\right)\Bigg|_{\omega=0}\,
\left(2M\right)^{\ell+2}\frac{(\ell+1)!}{(2\ell+1)!!}\,
(-1)^{\ell+1}\,
\Big[t-r_*-(t'-r_*')\Big]^{-(\ell+2)}\,.
\end{equation}
The leading-order contribution of $h\!\left(r'/2M-1\right)$ in the small-frequency expansion can be computed following~\cite{Rosato:2025rtr}, upon adopting the correct normalization and definition of the relevant functions, yielding
\begin{equation}
    h\!\left({r' \over 2M}-1\right)\Bigg|_{\omega=0}=\left({r' \over 2M}\right)^{\ell+1}+\mathcal{O}\left(\omega\right)\,,
\end{equation}
which gives
\begin{equation}
G^{\rm BC}_{\ell m}(r,r',t-t')
=2M\frac{(\ell+1)!(r')^{\ell+1}}{(2\ell+1)!!}\,
(-1)^{\ell+1}\,
\Big[t-r_*-(t'-r_*')\Big]^{-(\ell+2)}\,.
\end{equation}
This expression precisely reproduces the leading-order term of Eq.~\eqref{eq:timegreenfunct}, which is equivalent to Leaver's result~\cite{Leaver:1986gd}. It is therefore not surprising that the leading-order behavior coincides with that obtained for sources located outside the light ring. Indeed,~\cite{Rosato:2025rtr} derived the leading-order Green's function by performing an expansion around $M\omega=0$, which is insensitive to the source location (no far-zone approximation was employed).

More interestingly, the overall magnitude of the leading term scales as $(r'/M)^{\ell+1}$. This implies that, for sources located inside the light ring, the tail amplitude is suppressed relative to sources outside it. Such suppression has a natural physical interpretation: radiation generated inside the light ring must tunnel through the potential barrier before reaching infinity (and subsequently backscattering to produce tails). In the small-frequency regime, the tunneling probability (i.e., the greybody factor) is small, being suppressed by a factor $(M\omega)^{2\ell+1}$~\cite{Starobinskij2,Brito:2015oca}.

\subsection{QNMs, redshift terms, and horizon modes}\label{sec:redshift}
Very recently, considerable attention has been devoted to the role of near-horizon physics in black-hole ringdown signals. 
Beyond the standard ringdown, defined as superposition of QNMs, several works have identified additional structures associated with the near-horizon region: \emph{redshift terms}~\cite{DeAmicis:2025xuh} and  \emph{horizon modes} ~\cite{Mino:2008at,Zimmerman:2011dx,Laeuger:2025zgb}, both in agreement with mathematical-relativity estimates~\cite{Dafermos:2005eh}.

In the context of a Schwarzschild black hole, these contributions are characterized by purely negative imaginary frequencies related to integer multiples of the surface gravity.

More recently, it has been argued that such horizon modes are effectively screened for an observer at null infinity by greybody factors, so that they do not correspond to genuine propagating degrees of freedom in the asymptotic waveform~\cite{Oshita:2025qmn}. 
In this picture, although the near-horizon dynamics leaves an imprint in intermediate steps of the calculation, its contribution is suppressed in the standard scattering amplitude.

In this section we revisit this issue from the perspective of the analytic structure of the retarded Green's function in the frequency domain. 
First of all, we recover the result of Ref.~\cite{Oshita:2025qmn}. Indeed, if one considers the near-horizon expansion of the solution in Eqs.~\eqref{eq:nearhor} and~\eqref{eq:recurrenceHor}, it is clear that the $n$-th coefficient of the series expansion possesses a pole at
\begin{equation}
    \omega=\omega_H= -i{n \over 4M} \equiv -i n {\kappa_H}\,,
\end{equation}
with ${\kappa_H}$ denoting the surface gravity, this frequency exactly corresponding to the \emph{horizon mode} frequencies~\cite{Dafermos:2005eh,Mino:2008at,Zimmerman:2011dx,Laeuger:2025zgb,DeAmicis:2025xuh}
Despite the presence of these poles, however, when one considers the complete Green's function in Eq.~\eqref{eq:greenHor}, the term $1/A^{\rm in}_{\ell m \omega}$ is also present. Ref.~\cite{Zimmerman:2011dx} showed that $1/A^{\rm in}_{\ell m \omega}$ exactly vanishes at the horizon frequencies. Consequently, even though the solution possesses poles at those frequencies, the full Green's function does not. This is consistent with the results of Ref.~\cite{Oshita:2025qmn}, although derived in a different context.

However, inspired by the way these frequencies appear in Ref.~\cite{DeAmicis:2025xuh}, here we show how they emerge when considering the behavior of the Green's function in a neighborhood of the QNM poles, and how they survive at late times.

Consider a test particle plunging into a Schwarzschild black hole. 
In the time domain, the source can be written as in Eq.~\eqref{eq:testparticle_source}.

Near the horizon, the coefficient functions $f_{\ell m}$ and $g_{\ell m}$ are regular and vanish at least linearly in $(r-2M)$. 
To move to the frequency domain, we use the identity
\begin{equation}
\delta \bigl(r_*' - r_*'(t)\bigr)
=
\frac{\delta \bigl(t' - t_p(r')\bigr)}{\left|dr_*/dt'\right|}\,,
\end{equation}
where $t_p(r')$ is the coordinate time along the worldline, which admits the universal near-horizon form
\begin{equation}
t_p(r')=-\,r_*'+\sum_{k=0}^{\infty}c_k\,(r' - 2M)^k \,.
\end{equation}
In the above, a logarithmic divergence is contained in the term $-r_*'$, while the remainder is analytic in $(r'-2M)$. Therefore,
\begin{equation}
e^{i\omega t_p(r')}=e^{-i\omega r_*'}\left(1+\mathcal{O}(r'-2M)\right)\,,\quad {dr_* \over dt'}=-1+\sum_{k=1}^\infty t_k\,(r' - 2M)^k \,,
\end{equation}
It then follows that, in a neighborhood of the horizon, the Fourier-transformed source admits the expansion
\begin{equation}
\tilde{S}_{\ell m\omega}(r')
=
e^{-i\omega r_*'}
\sum_{n=1}^{\infty}
s^{\ell m}_n(\omega)\,
\left({r'\over 2M} - 1\right)^n \,.
\end{equation}

We now examine the effect of integrating over $r'$ in the convolution with the Green's function. First, we can write
\begin{equation}
\tilde{G}_{\ell m}(r_*,r'_*,\omega)\,\tilde{S}_{\ell m\omega}(r')
=
\frac{i\, e^{i\omega (r_* - 2r_*')}}{2\omega\, A^{\rm in}_{\ell m\omega}}\,
\sum_{n=1}^{\infty}
d^{\ell m}_n(\omega)\,
\left({r' \over 2M} - 1\right)^n\,,
\end{equation}
where the coefficients $d^{\ell m}_n(\omega)$ arise from multiplying the two separate series expansions in $r'/2M-1$. 

Next, we note that
\begin{equation}
 r'/2M-1=W\left(e^{-1+{r_*'/2M}}\right)\,,
\end{equation}
where $W(x)$ denotes the Lambert function, which admits a series expansion around $x=0$. 
Defining $x=e^{r_*/2M-1}$ and recalling that $r_*\to-\infty$ at the horizon, we obtain
\begin{equation}
    r'/2M-1=\sum_{n=0}^\infty \frac{(-n)^{n-1}}{n!\,e} e^{r_*/ 2M}\,,
\end{equation}
which converges for $r_*<0$. This allows us to express everything in terms of the tortoise coordinate in the region $r_*<0$
\begin{equation}
  G_{\ell m}(r_*,r'_*,\omega)S_{\ell m\omega}(r'_*)
=
\frac{i\, e^{i\omega r_*}}{2\omega\, A^{\rm in}_{\ell m\omega}}\,
\sum_{n=1}^{\infty}
\tilde{d}^{\ell m}_n(\omega)\,
e^{-2ir_*'\left( \omega + in{\kappa_H}\right)}\,. 
\end{equation}

We now consider the integral over $r_*'$. Care must be taken with the limits of integration:
\begin{itemize}
    \item The upper limit is $r_*=0$, where the series converges.
    \item The lower limit is determined by causality. As discussed in Ref.~\cite{DeAmicis:2025xuh}, the QNM signal emitted when the source is located at $r_*'$ and time $t_p(r')$ reaches the observer at time $t=t_p(r')+r_*+r_*'-4M\log\left(f(r')\right)$. Consequently, at time $t$ we observe only what was emitted at $2r'_*\sim r_*-t$, where we have used $t_p(r')\sim-r_*'$ and $r_*'-4M\log\left(f(r')\right)\sim -r_*'$ in the near-horizon limit of integration (we have fixed $t_0=0$ as the starting time).
\end{itemize} 
In the region $0<r_*<r_*(3M)$ we retain the previous expansion in $r/(2M)-1$.
We must evaluate
\begin{equation}
     \int_{(r_*-t)/2}^{r_*(3M)} dr_*'\,G_{\ell m}(r_*,r'_*,\omega)S_{\ell m\omega}(r'_*)
     =\frac{e^{i\omega r_*}}{4\omega\, A^{\rm in}_{\ell m\omega}}\left(
\sum_{n=1}^{\infty}
\frac{\tilde{d}^{\ell m}_n(\omega)}{\omega + in\kappa_H}\,
e^{-i\left(r_*-t\right)\left( \omega + in\kappa_H\right)}
-
\sum_{n=1}^{\infty}
d^{\ell m}_n(\omega)\,
p_n(\omega)e^{-i \omega t}\right)\,,
\label{eq:redshift_intermediate}
\end{equation}
where
\begin{equation}
   p_n(\omega)= \frac{e^{-6 i \omega }\, 2^{-n+4 i \omega }
   \left(1+n e^{2 i \omega }
   \mathrm{E}_{-n+4 i \omega +1}(2 i \omega )\right)}{\omega }\,,
\end{equation}
with $\mathrm{E}_\nu(z)$ denoting the generalized exponential integral.

Finally, moving back to the time-domain response and isolating the contribution of a single QNM pole at $\omega=\omega_q$, one obtains
\begin{equation}
    \Psi^{\rm QNMs}_{\ell m}(r_*,t)=\frac{1}{4\omega_q\, {dA^{\rm in}_{\ell m\omega}\over d\omega}\Big|_{\omega_q} }\,
\left(\sum_{n=1}^{\infty}
\frac{\tilde{d}^{\ell m}_n(\omega_q) e^{-n{\kappa_H} (t-r_*)}}{\omega_q + in{\kappa_H}}
-
\sum_{n=1}^{\infty}
d^{\ell m}_n(\omega_q)\,
p_n(\omega_q)e^{-i \omega_q (t-r_*)}\right)\,.
\end{equation}
The first sum gives rise to purely damped contributions of the form $e^{-n{\kappa_H} (t-r_*)}$ with $n>0$, which are associated with redshift terms and do not exhibit oscillatory behavior. 
The second sum instead retains the characteristic QNM time dependence $e^{-i\omega_q (t-r_*)}$ and therefore describes the standard QNM oscillation.
The coefficients $\tilde{d}^{\ell m}_n(\omega_q)$ and  $d^{\ell m}_n(\omega_q)$ depend on the specific source and encode the near-horizon structure of the excitation.
The \textit{redshift terms}, as defined here and in Ref.~\cite{DeAmicis:2025xuh}, are a near-horizon contribution of the Green's function poles at the \textit{quasinormal frequencies}, which emerges once causality is accounted for. In particular, it is required that signals emitted by the horizon-approaching source travel on the light cone
(see the lower limit of integration in Eq.~\eqref{eq:redshift_intermediate}).
The \textit{horizon modes} as defined in Refs.~\cite{Mino:2008at,Zimmerman:2011dx} arise from a different spectral component: they are additional poles of the convolution integral between the source and the Green's function in the frequency domain, at the so-called \textit{horizon frequencies}.
Ref.~\cite{Oshita:2025qmn} showed that the horizon modes are screened by the potential barrier peak and do not propagate to $\mathcal{I}^+$, due to other components of the Green's function vanishing at the aforementioned horizon frequencies.
The redshift terms, instead, are not canceled by other features of the Green's function (e.g. the branch cut, as shown in the previous section), and persist up to late times. 
The physical interpretation is that an observer at $\mathcal{I}^+$ never ``sees'' the test-particle crossing the horizon through gravitational radiation.

\subsubsection{Numerical indication of a redshift contribution}
\begin{figure}[t]
\centering
\includegraphics[width=\textwidth]{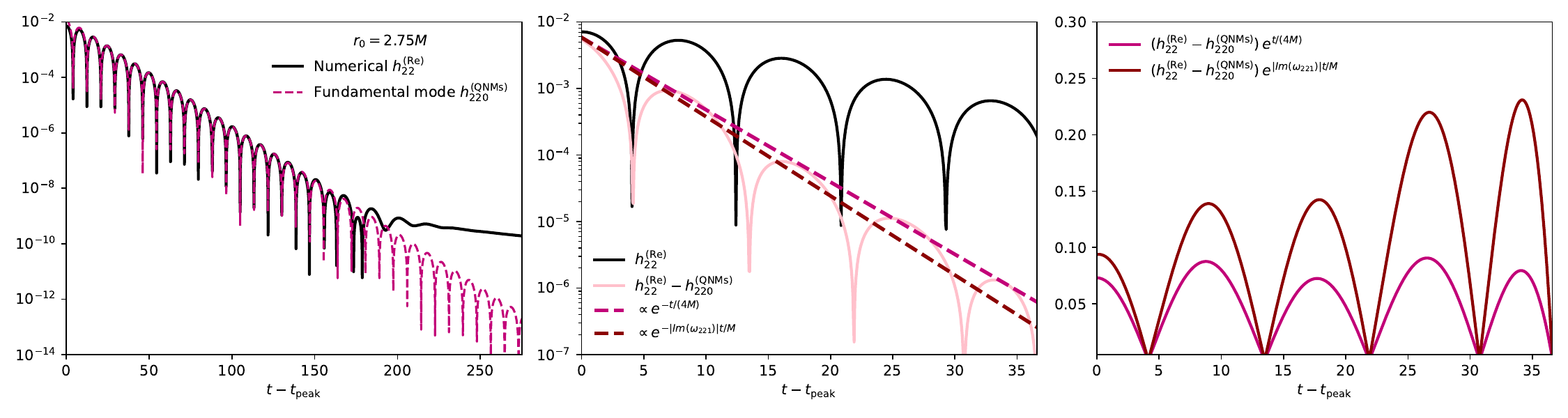}
\caption{
Representative examples suggesting the presence of a redshift contribution in the ringdown signal. 
\textit{Left:} real part of the waveform $h_{22}$ produced by a test particle in radial infall starting from $r_0=2.75M$, together with the fundamental QNM contribution $h_{220}^{\rm (QNMs)}$. The amplitude and phase of the fundamental mode are obtained from a fit performed for several starting times $t_{\rm start}$; the value shown corresponds to a region where the fitted parameters become approximately stable (around $t_{\rm start}\gtrsim 40M$). 
\textit{Middle:} residual signal $h_{22}-h_{220}^{\rm (QNMs)}$ compared with two reference decays, $e^{-t/(4M)}$ (the first Schwarzschild redshift term) and $e^{-|\Im(\omega_{221})|t/M}$ (the first overtone). The residual appears qualitatively more consistent with the slower decay associated with the redshift term. 
\textit{Right:} rescaled residuals $( {\rm Re}\,h_{22}-h_{220}^{\rm (QNMs)} )e^{t/(4M)}$ and $( {\rm Re}\,h_{22}-h_{220}^{\rm (QNMs)} )e^{|\Im(\omega_{221})|t/M}$. The latter remains approximately constant only during the first two oscillations, whereas at later times the decay becomes slower, suggesting a behavior closer to that expected from the redshift contribution.
}
\label{fig:redshift}
\end{figure}

To complement the analytic discussion, we present in Fig.~\ref{fig:redshift} a numerical experiment of a radial plunge starting inside the light ring, at $r_0=2.75\,M$, where a redshift contribution is expected to be present. Although this analysis does not allow for an unambiguous extraction of such a term, it provides a suggestive qualitative indication in that direction.

The left panel of Fig.~\ref{fig:redshift} shows the numerical waveform $h_{22}$ together with the fundamental-mode contribution ${h}_{220}^{\rm (QNMs)}$. The amplitude and phase of the fundamental mode are obtained from a fit of the form $A e^{-i\omega_{220} t+i\phi}$ performed in a time window where the fitted parameters become approximately stable (around $t_{\rm start}\gtrsim 40M$). As expected, the fundamental mode provides a good description of the post-peak signal, but a non-negligible residual remains. 

In the central panel we compare the decay of this residual, $h_{22}-{h}_{220}^{\rm (QNMs)}$, with two reference behaviors: $e^{-t/(4M)}$, corresponding to the first Schwarzschild redshift term, and $e^{-|\Im(\omega_{221})|t} \simeq e^{-0.274\,t/M}$, corresponding to the first overtone. The residual appears more compatible with the slower decay rate $e^{-t/(4M)}$ than with the overtone decay, especially beyond the earliest oscillations.

This trend is further illustrated in the right panel of Fig.~\ref{fig:redshift}, where we plot 
\[
(\,h_{22}-{h}_{220}^{\rm (QNMs)})\,e^{t/(4M)}
\qquad \text{and} \qquad
(\,h_{22}-{h}_{220}^{\rm (QNMs)})\,e^{|{\rm Im}(\omega_{221})|t/M}.
\]
If the residual were dominated by the first overtone, the second combination would remain approximately constant. Instead, this behavior is realized only during the first two oscillations after the peak, whereas at later times the rescaled signal drifts significantly. By contrast, the combination multiplied by $e^{t/(4M)}$ appears overall more stable, suggesting that the actual decay rate is slower and closer to that of the first redshift contribution.

A clean quantitative disentangling remains difficult, since the first overtone and the first redshift term have comparable amplitudes and rather similar damping times, making standard fits highly degenerate. Nevertheless, the numerical behavior suggests the following qualitative picture: up to roughly $20M$ after the peak, the first overtone provides an effective description of the residual, whereas at slightly later times the slower decay associated with the redshift term may become dominant. In this sense, the numerical data appear consistent with the analytic expectation that a redshift contribution can persist at later times relative to the first overtone.

Finally, we confirm that the tail contribution is strongly suppressed for sources located inside the light ring, with an amplitude significantly smaller than in the case of infalls starting outside the light ring.

\subsection{Greybody-factor description inside the light ring}\label{sec:gfsinsidelightring}
As discussed above, the decomposition of the retarded Green's function
\begin{equation}
G_{\ell m} = G^{(1)}_{\ell m} + G^{(2)}_{\ell m}
\end{equation}
remains formally valid also in the region $r' < 3M$.
However, inside the light ring the intuitive interpretation of 
$G^{(1)}_{\ell m}$ as a direct response and of $G^{(2)}_{\ell m}$ as a
reflected contribution generated by the effective potential becomes less transparent.

To clarify this point, it is useful to revisit the one-dimensional scattering toy model analyzed in Sec.~\ref{sec:greybody}, now considering a rectangular barrier
\begin{equation}
V(x) =
\begin{cases}
h \,, & x \in [-x_0/2,\, x_0/2] \\
0 \,, & \text{otherwise}
\end{cases}\,\,.
\end{equation}
We consider a source localized to the left of the barrier ($x < -x_0/2$), 
and an observer at $x \to +\infty$, imposing purely incoming (left-moving) boundary conditions at $x \to -\infty$.
This setup mimics the physical boundary conditions of a black hole spacetime.

In this case, the Green's function at infinity takes the schematic form
\begin{equation}
G(\omega) =\frac{i\,e^{\,i\omega(x-x')}}{2\omega A^{\rm in}(\omega)}\,,
\end{equation}
where $1/A_{\rm in}(\omega)$ is the transmission amplitude across the barrier.
The physical interpretation is straightforward: only the component of the source that is initially right-moving can reach the observer, and it must tunnel through the potential barrier, leading to the suppression factor $1/A^{\rm in}(\omega)$.

The analogous situation in the black hole case for sources located inside the light ring leads to a Green's function of the same structure,
\begin{equation}
\tilde{G}_{\ell m}(r_*,r'_*,\omega)=\frac{i\, e^{i\omega (r_* - r_*')}}{2\omega\, A^{\rm in}_{\ell m\omega}}\,
h\!\left(\frac{r'}{2M}-1\right)\,,
\end{equation}
with $h\left(r’/(2M)-1\right)$ defined at the beginning of Sec.~\ref{sec:sourceinside}.
As discussed, this configuration of the Green's function does not exhibit a pole at $\omega = 0$, and therefore the prompt response is absent.
A power-law tail is still present, but it is strongly suppressed, since the relevant radiation must tunnel through the potential barrier.
As a consequence, the signal is dominated by the contributions associated with the poles of $A^{\rm in}_{\ell m}(\omega)$, i.e. the quasinormal frequencies.
These include both the standard quasinormal mode oscillations and the redshifted  terms discussed in Sec.~\ref{sec:redshift}.

How does this picture connect with the decomposition  $G_{\ell m} = G^{(1)}_{\ell m} + G^{(2)}_{\ell m}$?
For $r' < 3M$, the prompt response is no longer clearly identifiable, since the $\omega = 0$ pole contributions associated with $G^{(1)}_{\ell m}$ and $G^{(2)}_{\ell m}$ cancel in this region. Equivalently, the prompt response carried by $G^{(1)}_{\ell m}$ becomes effectively instantaneous, i.e. it is supported over a vanishingly small time interval.
In particular, there is no longer a time delay separating the signals carried by $G^{(1)}_{\ell m}$ and $G^{(2)}_{\ell m}$.
Moreover, although a tail contribution persists, its amplitude is strongly suppressed by transmission through the potential barrier. This implies that the pole contributions at the quasinormal frequencies dominate the signal in this framework.

Importantly, the decomposition $G_{\ell m} = G^{(1)}_{\ell m} + G^{(2)}_{\ell m}$ is unaffected by the location of the source. In particular, the quasinormal poles $\omega_q$ arise solely from the zeros of $A^{\rm in}_{\ell m}(\omega)$ and therefore reside entirely in $G^{(2)}_{\ell m}$, both outside and inside the light ring. Consequently, although the physical interpretation differs from that discussed in Sec.~\ref{sec:greybody}, the dynamics remains entirely encoded in $G^{(2)}_{\ell m}$ even for sources inside the light ring, ensuring that the model of Eq.~\eqref{eq:greybodyRD} remains applicable.

Therefore, even for sources located inside the light ring, the ringdown signal observed at infinity is always proportional to $R_{\ell m}(\omega)$.
While the physical interpretation of the intermediate-time response becomes more subtle, the frequency-domain structure of the signal remains unchanged.
\section{Conclusion}
In this work we revisited the analytic structure of the Schwarzschild retarded Green's function and provided a unified interpretation of its causal decomposition in the frequency domain.

Previous analyses~\cite{DeAmicis:2025xuh, DeAmicis:2026tus} have shown that the decomposition of the Green's function into $G^{(1)}_{\ell m}$ and $G^{(2)}_{\ell m}$ (see Eq.~\eqref{eq:G1G2}) has a clear causal interpretation at the level of the prompt response and of the QNM poles. Here we have extended this structure to the branch-cut contribution, demonstrating that the late-time tails associated with $G^{(1)}_{\ell m}$ and $G^{(2)}_{\ell m}$ are likewise causally disconnected. In particular, the tail generated by $G^{(2)}_{\ell m}$ reaches the observer after a well-defined delay with respect to that generated by $G^{(1)}_{\ell m}$, completing the causal picture for all singular structures entering the Green's function.

By performing a systematic small-frequency expansion beyond leading order, we derived the full hierarchy of subleading corrections to the late-time tail, showing that logarithmic terms arise unavoidably at each order due to the branch-cut structure. Numerical comparisons confirm that these corrections are quantitatively relevant when reconstructing the tail at intermediate times, in the presence of non-negligible ringdown contributions. Thus, subleading tail effects might compete with higher-order overtones. As an extension of our work, it would be interesting to understand their interplays in black-hole merger signals.

We further showed that this causal organization has a transparent interpretation in terms of greybody factors. In the frequency domain, the portion of the signal associated with the quasinormal ringing and with the dominant late-time behavior is entirely controlled by the reflection coefficient of the black hole. This provides the first theoretical foundation for recent frequency-domain phenomenological models in which the ringdown signal is described as a greybody-modulated response~\cite{Oshita:2023cjz,Rosato:2024arw,Okabayashi:2024qbz,Rosato:2025ulx}.

For sources located inside the light ring, the separation into direct and reflected components loses its simple geometric interpretation, since the notion of reflection off the potential barrier is no longer defined. Nevertheless, the physically relevant portion of the signal remains controlled by the component associated with $G^{(2)}_{\ell m}$, which contains the reflection coefficient. In this sense, the greybody-factor description of the ringdown persists even in the region $r' < 3M$, although the causal hierarchy between direct and reflected channels no longer admits a transparent space-time picture. 

In this regime, we also clarify the role of redshift terms. These contributions do arise when the source is located inside the light ring. Contrary to what discussed in previous work~\cite{Mino:2008at,Zimmerman:2011dx,Oshita:2025qmn}, they do not correspond to independent pole structures (screened at late times) of the frequency-domain convolution integral between a test-particle source and the Green's function. Rather, they emerge as a direct consequence of the causal implementation of boundary conditions in the frequency domain, which effectively redshifts the QNM contributions. The redshift terms are therefore not new singularities of the Green's function, but modified manifestations of the same QNM structure once causality is properly enforced for near-horizon sources.

Overall, our results clarify the relation between the analytic structure of the Green's function, its causal decomposition in the time domain, and the role of greybody factors in perturbation theory, providing both conceptual insight and guidance for analytic modeling of black-hole ringdown signals.

A natural extension is to consider the Green's function of a spinning black hole. This could be achieved by performing a Detweiler transformation~\cite{Detweiler:1977gy,Maggio:2018ivz}, which maps the Teukolsky equation governing perturbations of a Kerr black hole~\cite{Teukolsky:1973ha} to the same form as in Eq.~\eqref{Schroedinger} with a real potential. 

\begin{acknowledgements}
 RFR acknowledges the hospitality of Perimeter Institute for Theoretical Physics, where part of this work was completed during a research visit. This work is partially supported by the MUR FIS2 Advanced Grant ET-NOW (CUP:~B53C25001080001) and by the INFN TEONGRAV initiative.
This research was supported in part by
Perimeter Institute for Theoretical Physics. Research at Perimeter Institute is supported in part by the Government of Canada through the Department of Innovation, Science and Economic Development and by the Province of Ontario through the Ministry of Colleges and Universities.   
\end{acknowledgements}

\newpage
\appendix
\section{Source}\label{app:source}
We list the even-sector source functions $f_{\ell m}(t,r_*),\,g_{\ell m}(t,r_*)$ introduced in Eq.~\eqref{eq:testparticle_source}, as defined in Ref.~\cite{Nagar:2006xv}(see also~\cite{Martel:2003jj,Sasaki:2003xr,Martel:2005ir})
\begin{align}
f_{\ell m}(t,r_*)\equiv & -\frac{16\pi\mu Y_{\ell m}^*}{r\hat{H}\lambda \left[r(\lambda-2)+6\right]}\left(1-\frac{2M}{r}\right) \,\left\lbrace -2i m \,p_{\varphi}p_{r_*} + 5+\frac{12\hat{H}^2 r}{r(\lambda-2)+6}-\frac{r\lambda}{2}+\frac{2p^2_{\varphi}}{r^2}\right.\notag \\
&\left. +\frac{p_{\varphi}^2}{r^2(\lambda-2)}\left[r(\lambda-2)(m^2-\lambda-1)+2(3m^2-\lambda-5)\right] \right\rbrace \, ,\\ 
g_{\ell m}(t,r_*)\equiv& -\frac{16\pi\mu Y_{\ell m}^*}{r\hat{H}\lambda \left[r(\lambda-2)+6\right]} \left(1-\frac{2M}{r}\right) \,(p^2_{\varphi} + r^2) \, ,
\end{align}
where $\lambda\equiv\ell(\ell+1)$ and $\mu$ is the mass of the test-particle. The $\mu$-rescaled energy, $\hat{H}$ is defined as
\begin{equation}
\hat{H}=\sqrt{A\left(1+\frac{p_{\varphi}^2}{r^2}\right)+p_{r_*}^2} \, .
\end{equation}
We focus on equatorial trajectories, so the polar coordinate of the test-particle is $\theta_0=0$. Its azimuthal coordinate is $\varphi$ and $(p_r,p_{\varphi})$ are the $\mu$-rescaled momenta associated to the variables $(r,\varphi)$. The evolution of these quantities along the trajectory is computed by solving the Hamiltonian equations of motion
\begin{equation}
\begin{split}
    &\dot{r}=\frac{A}{\hat{H}}p_{r_*} \, ,\\
    &\dot{\varphi}=\frac{A}{r^2\hat{H}}p_{\varphi} ,\\
    &\dot{p}_{r_*}=A\hat{\mathcal{F}}_{r}-\frac{A}{r^2\hat{H}}\left(p^2_{\varphi}\frac{3-r}{r^2}+1 \right) \, ,\\
    &\dot{p}_{\varphi}=\hat{\mathcal{F}}_{\varphi} \, .
\end{split}
\end{equation}
Note that these equations are driven by radiation-reaction effective forces, more details are given in Refs.~\cite{Chiaramello:2020ehz,Albanesi:2021rby}.

\section{Recurrence relation}\label{app:recurrence}
We expressed the homogeneous solutions to Eq.~\eqref{eq:Regge-Wheeler} as a series of Coulomb wave functions in Eq.\eqref{eq:psirhmin}, whose coefficients are the solution of the recurrence relation 
\begin{equation}\label{eq:recurrence}
\alpha_n a_{n+1}+\beta_n a_{n}+\gamma_n a_{n-1}=0\,.
\end{equation}
The coefficient of previous relation are defined as 
\begin{align}\label{eq:coeffrecurrence}
\alpha_n &= 
-\frac{(-2Mi\omega) R_{n+1}}{2n + 2\nu + 3}
\Big[
(n+\nu+1)(n+\nu+2)
-(n+\nu+2)\big(2(-2Mi\omega)+1\big)
-s^2+((-2Mi\omega)+1)^2
\Big], \notag\\[6pt]
\beta_n &= 
(n+\nu)(n+\nu+1)
+3(-2Mi\omega)^2 - \ell(\ell+1)
+(-2Mi\omega) Q_n \big[(n+\nu)(n+\nu+1)-s^2+(-2Mi\omega)^2\big], \\[6pt]
\gamma_n &= 
-\frac{(-2Mi\omega) R_n}{2n + 2\nu - 1}
\Big[
(n+\nu)(n+\nu-1)
+(n+\nu-1)\big(2(-2Mi\omega)+1\big)
-s^2+((-2Mi\omega)+1)^2
\Big],\notag
\end{align}
where
\begin{equation}\label{eq:QR}
Q_n = \frac{\eta}{(n+\nu)(n+\nu+1)}, 
\qquad
R_n = \frac{\sqrt{(n+\nu)^2 + \eta^2}}{n+\nu}\,.
\end{equation}
The recurrence relation can be solved by fixing the initial condition $a_0 = 1$. 
In the small-frequency limit, one finds that $a_n \sim \omega^{|n|}$ at leading order~\cite{Casals:2015nja}, 
so that the coefficients $a_n$ and the renormalized angular momentum $\nu$ 
can be determined iteratively as power series in $\omega$. 
Following the standard procedure~\cite{Casals:2015nja}, we expand $\nu$ as
\begin{equation}\label{eq:nu}
\nu = \ell + \sum_{j=1}^{\lfloor N_{\max}/2 \rfloor} \nu_{2j}\,(M\omega)^{2j},
\end{equation}
where only even powers of $\omega$ appear in the expansion. 
The recurrence relation is then solved order by order in $\omega$, 
yielding a coupled system of algebraic equations for the coefficients $\{a_n,\,\nu_{2j}\}$. 
At each order, the equations are linear in the unknowns and can be solved sequentially, 
starting from $a_0 = 1$. 
In the limit $\omega \to 0$, the renormalized angular momentum $\nu$ admits two possible asymptotic values, 
namely $\nu \to \ell$ and $\nu \to -\ell-1$, corresponding to the two linearly independent Coulomb-type solutions. 
Throughout this work, we select the branch $\nu \to \ell$, 
which ensures regularity of the solution at the horizon and continuity with the standard 
Regge–Wheeler solution in the Schwarzschild limit.
As a matter of example here we list the result up to the order $(M\omega)^2$, similar results can be obtained at all orders.
\begin{align}
&a_0=1\,,\\
&a_1=\frac{M \omega \left[(1+\ell)\left(1+\ell-4 i M \omega\right)-s^{2}\right]}
{(1+\ell)(1+2\ell)}-\frac{4 i(M \omega)^2}
{(1+2\ell)}+\mathcal{O}\left((M\omega)^3\right)\,,
\\
&a_2=\frac{M^{2}\omega^{2}\,(1+\ell-s)(2+\ell-s)(1+\ell+s)(2+\ell+s)}
{(1+\ell)(1+2\ell)(3+2\ell)^{2}}+\mathcal{O}\left((M\omega)^3\right)\,,
\\
&
a_{-1}=\frac{M \omega \left[-\ell^2+s^{2}\right]}
{\ell(1+2\ell)}-\frac{4i(M\omega)^2}{1+2\ell}+\mathcal{O}\left((M\omega)^3\right)\,,
\\
&
a_{-2}=\frac{M^{2}\omega^{2}\,(\ell-s)(\ell+s)\left[(\ell-1)^{2}-s^{2}\right]}
{(1-2\ell)^{2}\,\ell(1+2\ell)}+\mathcal{O}\left((M\omega)^3\right)\,,\\
&\nu_2=\frac{2 \left(-6 \left(\ell^2+\ell-1\right) s^2+\ell (\ell+1) (33 \ell (\ell+1)-25)-3 s^4\right)}{\ell (\ell+1) (2 \ell-1) (2 \ell+1) (2 \ell+3)}\,.
\end{align}

\section{Time dependence of late time tail}\label{app:timedependencetail}

In order to compute the time-domain Green's function, one needs to evaluate the
frequency integral appearing in Eq.s~\eqref{eq:greenintegral} and ~\eqref{eq:greenintegraleven}. In particular, we consider
the contribution from the branch cut along the negative imaginary axis.
Throughout this appendix we assume $n\in\mathbb{N}_0$ and $k\in\mathbb{N}_0$.
We therefore define
\begin{equation}
I_{n+\ell,k}(\tau)
=
\int_{-i\infty}^{0}
(M\omega)^{n+\ell} \,[\log(M\omega)]^{k}\,
e^{-i\omega\tau }\, d\omega ,
\qquad\tau >0 .
\end{equation}

The integral is evaluated by deforming the contour along the negative imaginary
axis and choosing the principal branch of the logarithm, $\arg\omega=-\pi/2$.
The resulting integral reduces to a Laplace transform and can be computed in
closed form.
For generic $n,k\in\mathbb{N}_0$, one finds
\begin{equation}
I_{n+\ell,k}(\Delta t)
=
(-i)^{\,n+1} M^n\,\frac{n!}{(\tau)^{n+\ell+1}}
\sum_{j=0}^{k}
\binom{k}{j}
\Bigl(\log \left({M \over i\tau}\right)\Bigr)^{k-j}
\mathcal{P}_{j}(n+\ell),
\end{equation}
where $\mathcal{P}_{j}(n+\ell)$ are polynomials in $n+\ell$ generated by derivatives of
$\Gamma(n+\ell+1)$
\begin{equation}
\mathcal{P}_{j}(n+\ell)
\equiv
\left.
\frac{1}{\Gamma(n+\ell+1)}
\left(\frac{d^{\,j}}{d\alpha^{\,j}}
\Gamma(\alpha+1)\right)
\right|_{\alpha=n+\ell}\,,
\end{equation}
which can be expressed in terms of harmonic numbers. For example, $\mathcal{P}_{0}(n+\ell)=1$ and $\mathcal{P}_{1}(n+\ell)=H_{n+\ell}-\gamma_E$, with
$H_{n+\ell}=\sum_{j=1}^{n+\ell} j^{-1}$.
Consequently, the first two cases read
\begin{align}
I_{n+\ell,0}(\tau)
&=
(-i)^{\,n+\ell+1} M^n\,\frac{n+\ell!}{(\tau)^{n+\ell+1}},
\\[1ex]
I_{n+\ell,1}(\tau)
&=
\,(-i)^{\,n+\ell+1} M^n\,\frac{n+\ell!}{(\tau)^{n+\ell+1}}
\Bigl[H_{n+\ell}-\gamma+\log \left({M \over i \tau}\right)\Bigr].
\end{align}

In general, the integral $I_{n+\ell,k}$ produces a late-time behavior proportional to
$\tau^{-(n+\ell+1)}$ multiplied by powers of $\log(\tau)$ up to order $k$. Notice that being always $k\leq n+\ell$, the terms $\tau^{-(n+\ell+1)}\log(\tau/M)^k$ are always convergent for $\tau\gg M$.
\section{Series expansion of the Green's  function}\label{app:coeffGreen}
For completeness, we report here the coefficients entering Eq.~\eqref{eq:greenintegral} 
up to order $N_{\rm max}=3$ for the $\ell=-s=2$ perturbation. For the odd sector we find
\begin{align}
     \tilde{g}^{(0 0),\,o}_{21}(r') &=-\frac{4 \left(r'\right)^3}{5}\,, \notag\\ 
     \tilde{g}^{(1 0),\,o}_{21}(r') &=\frac{8}{3045}\left(-2436 \left(r'\right)^4+667 \left(r'\right)^3+4872 \left(r'\right)^2+168 r'+6496\right)\,,\notag \\ 
     \tilde{g}^{(1 1),\,o}_{21}(r') &=\frac{32}{5} \left(r'\right)^3\notag \\ 
     \tilde{g}^{(2 0),\,o}_{21}(r') &=-\frac {8 } {63945}\left(292320 \left(r' \right)^5 + 
   126672 \left(r' \right)^4 + 490896 \left (r' \right)^2 - 
   1361556 r'-145 (10985 + 2436 \log(2)) \left(r' \right)^3 + 100688 \right)\,,\notag \\ 
     \tilde{g}^{(2 1),\,o}_{21}(r') &=64 \left(r'\right)^4-\frac{1712 \left(r'\right)^3}{35}-128 \left(r'\right)^2-\frac{128 r'}{29}-\frac{512}{3}\,, \notag\\ 
     \tilde{g}^{(2 2),\,o}_{21}(r') &=-32 \left(r'\right)^3\,.
     \notag
\end{align}
For the even sector we find
\begin{align}
         \tilde{g}^{(0 0),\,e}_{22}(r') &=-\frac{r' \left(4 \left(r'\right)^3+4 \left(r'\right)^2+r'-2\right)}{5 \left(r'+1\right)}\,, \notag\\ 
         \tilde{g}^{(1 0),\,e}_{22}(r') &=\frac{-19488 \left(r'\right)^7-16588 \left(r'\right)^6+51620 \left(r'\right)^5+66623 \left(r'\right)^4+40354 \left(r'\right)^3+51632 \left(r'\right)^2+49952
   r'-25984}{3045 \left(r'\right)^2 \left(r'+1\right)}\,,\notag \\ 
       \tilde{g}^{(1 1),\,e}_{22}(r') &=\frac{8 r' \left(4 \left(r'\right)^3+4 \left(r'\right)^2+r'-2\right)}{5 \left(r'+1\right)}\,,\notag \\ 
         \tilde{g}^{(2 0),\,e}_{22}(r') &=\frac{1}{255780 \left(r'\right)^2 \left(r'+1\right)}\left(-9354240 \left(r'\right)^8-16038624 \left(r'\right)^7+17874416 \left(r'\right)^2-62908384 r'+4339328\right.\notag\\&\left.+116 (469283+97440 \log (2)) \left(r'\right)^6+4 (14076583+2825760 \log
   (2)) \left(r'\right)^5+(5666377+2825760 \log (2)) \left(r'\right)^4\right.\notag\\&\left.+(9259566-5651520 \log (2)) \left(r'\right)^3+\right)\,,\notag \\ 
        \tilde{g}^{(21),\,e}_{22}(r')
        &=\frac{2}{3045 \left(r'\right)^2 \left(r'+1\right)} \left(97440 \left(r'\right)^7+35148 \left(r'\right)^6-305892 \left(r'\right)^5-345063 \left(r'\right)^4-177874 \left(r'\right)^3-258160 \left(r'\right)^2\right.\notag\\&\left.-249760
   r'+129920\right)\,,\notag\\
         \tilde{g}^{(2 2),\,e}_{22}(r') &=-\frac{8 r' \left(4 \left(r'\right)^3+4 \left(r'\right)^2+r'-2\right)}{r'+1}\,. \notag
\end{align}

with the $\ell=2\,,m=0$ multipole coinciding with the $\ell=m=2$ one. 
\newpage
\bibliography{biblio}
\end{document}